\documentclass[preprint2, tighten]{aastex63} 
\usepackage{amsmath, amssymb, latexsym}
\usepackage{bigints}
\usepackage{savesym}
\savesymbol{tablenum}
\usepackage{siunitx}
\restoresymbol{SIX}{tablenum}
\usepackage{hyphenat}
\usepackage{graphicx}
\usepackage{float}
\usepackage{ulem}
\usepackage{mathtools}
\usepackage{comment}
\usepackage{microtype}
\usepackage{longtable}

\newcommand{\change}[1]{\textcolor{black}{#1}}
\newcommand{\update}[1]{\textcolor{black}{#1}}
\newcommand{\maybe}[1]{\textcolor{black}{#1}}

\begin{document}

\title{Primordial Black Hole Dark Matter Simulations Using PopSyCLE}

\email{pruett6@llnl.gov}

\author[0000-0002-2911-8657]{Kerianne Pruett}
\affiliation{Lawrence Livermore National Laboratory, Livermore, CA 94550}

\author[0000-0003-0248-6123]{William Dawson}
\affiliation{Lawrence Livermore National Laboratory, Livermore, CA 94550}

\author[0000-0002-7226-0659]{Michael S. Medford}
\affiliation{Department of Astronomy, University of California, Berkeley, CA 94720}
\affiliation{Lawrence Berkeley National Laboratory, Berkeley, CA 94720}

\author[0000-0001-9611-0009]{Jessica R. Lu}
\affiliation{Department of Astronomy, University of California, Berkeley, CA 94720}

\author[0000-0002-6406-1924]{Casey Lam}
\affiliation{Department of Astronomy, University of California, Berkeley, CA 94720}

\author[0000-0002-5910-3114]{Scott Perkins}
\affiliation{Lawrence Livermore National Laboratory, Livermore, CA 94550}

\author[0000-0002-1052-6749]{Peter McGill}
\affiliation{Lawrence Livermore National Laboratory, Livermore, CA 94550}

\author[0000-0003-2632-572X]{Nathan Golovich}
\affiliation{Lawrence Livermore National Laboratory, Livermore, CA 94550}

\author{George Chapline}
\affiliation{Lawrence Livermore National Laboratory, Livermore, CA 94550}

\begin{abstract}

Primordial black holes (PBHs), theorized to have originated in the early universe, are speculated to be a viable form of dark matter. If they exist, they should be detectable through photometric and astrometric signals resulting from gravitational microlensing of stars in the Milky Way. Population Synthesis for Compact\hyp{}object Lensing Events, or \texttt{PopSyCLE}, is a simulation code that enables users to simulate microlensing surveys, and is the first of its kind to include both photometric and astrometric microlensing effects, which are important for potential PBH detection and characterization. To estimate the number of observable PBH microlensing events we modify \texttt{PopSyCLE} to include a dark matter halo consisting of PBHs. We detail our PBH population model, and demonstrate our \texttt{PopSyCLE} + PBH results through simulations of the OGLE\hyp{}IV and Roman microlensing surveys. We provide a proof\hyp{}of\hyp{}concept analysis for adding PBHs into \texttt{PopSyCLE}, and thus include many simplifying assumptions, such as $f_{\text{DM}}$, the fraction of dark matter composed of PBHs, and $\bar{m}_{\text{PBH}}$, mean PBH mass. Assuming $\bar{m}_{\text{PBH}}=30$ $M_{\odot}$, we find $\sim$ \maybe{3.6}$f_{\text{DM}}$ times as many PBH microlensing events than stellar evolved black hole events, a PBH average peak Einstein crossing time of $\sim$ \maybe{91.5} days, \change{estimate on order of $10^2f_{\text{DM}}$ PBH events within the 8 year OGLE-IV results}, and estimate Roman to detect $\color{black}\sim$ \update{1,000}$f_{\text{DM}}$ PBH microlensing events throughout its planned microlensing survey.

\end{abstract}


\section{Introduction}
\label{sec:Intro}

Approximately 85\% of the matter in our universe is composed of dark matter, which is not well\hyp{}understood. Dark matter cannot be observed via electromagnetic mechanisms (hence ``dark") and is only observable through gravitational interactions, making it challenging to detect. Primordial black holes (PBHs), hypothesized in the 1960s \citep{zel:1967}, are a theoretical type of black hole believed to be a viable dark matter candidate. Unlike black holes that are formed through stellar evolution, PBHs are thought to have formed in the early universe \citep[within 1 second of the Big Bang, e.g.,][]{Carr:2018}, via gravitational collapse in over\hyp{}dense regions. PBHs are theorized to have formed during the radiation dominated era, before Big Bang nucleosynthesis, thus they are considered non\hyp{}baryonic and are speculated to behave like ``cold" dark matter. Our current understanding of the universe places a $5\%$ constraint on the baryonic matter fraction \citep{Planck:2014}, meaning that stellar evolved black holes can never account for more than $5\%$ of the energy density in our universe. Thus, while we know that stellar evolved black holes exist and may contribute to the dark matter fraction, we know they cannot explain the majority of the missing dark matter in our universe. 

Because PBHs formed at roughly the cosmological horizon \citep{Chapline:1975a}, their initial masses should correspond to the horizon mass:

\begin{equation*}
    M \sim \frac{c^3 t}{G},
\end{equation*}
where $c$ is the speed of light, $t$ is the time since the big bang, and $G$ is the gravitational constant \citep{Carr:1975, Carr:2021}. This means that PBHs should have masses between $\sim$$10^{-8}$ $kg$ (if formed at Planck time $t = 10^{-43}$ $s$) and $\sim$$10^5$ $M_{\odot}$ (if formed at $t = 1$ $s$) \citep{Carr:1975}. PBHs are considered ``free\hyp{}floating" in the late universe, and should retain roughly their initial masses \citep{Chapline:1975a}. The initial conjecture of PBHs drove \citet{Hawking:1971} to study quantum gravitational effects of black holes, and led to the discovery that PBHs formed before $t=10^{-23}$ $s$ ($M \sim 10^{12}$ $kg$) should have evaporated by present day \citep{Hawking:1974}. 

In 2016, the Laser Interferometer Gravitational\hyp{}Wave Observatory (LIGO) detected gravitational waves from two merging $\sim$30 $M_{\odot}$ black holes \citep{Abbott:2016}, which were potentially primordial in origin \citep{Bird:2016}. The idea of PBHs as a viable form of dark matter had nearly died out over the last few decades due to insufficient new evidence or microlensing results \citep[e.g.,][etc.]{Alcock:1993, Aubourg:1993, Croon:2020, Udalski:1994, Wyrzykowski:2011}, however, our inability to detect dark matter (of any variety) throughout that time, combined with recent scientific discoveries, has reinvigorated scientists to explore PBHs as a potential dark matter candidate. These recent explorations have placed very tight constraints on $f_{\text{DM}}$ covering many decades in mass \citep[see summary in][]{Bird:2022}.

If PBHs exist, they should be present within our Milky Way galaxy \citep{Chapline:1975b} and observable via gravitational microlensing signals from local stars \citep[e.g.,][]{Chapline:2016}. Microlensing is the effect when a foreground object (the lens) passes in front of a luminous background object (the source), relative to the observer, in such a way that causes magnification of the source flux. While microlensing can cause magnification on the same order as strong gravitational lensing (i.e. multiple images), it happens on a smaller angular scale, such that lensed images are too small to be resolved. 

Microlensing imparts a characteristic photometric signal on the measured light curves of lensed stars, which can be used to detect and characterize foreground objects irrespective of their luminosity, or whether they are gravitationally bound to a companion. Unlike other probes, such as gravitational waves or X\hyp{}ray binaries, microlensing can detect free\hyp{}floating primordial black holes, in addition to binary black hole systems, and enables us to infer properties of the lens without observing the lens itself. Because primordial black holes interact gravitationally and not electromagnetically, microlensing is the most direct means of probing PBHs.

The angular Einstein radius of a microlensing event, $\theta_E$, sets the angular scale of the microlensing event, and is given by:

\begin{equation}
    \theta_E = \sqrt{\frac{4GM}{c^2} \Bigg( \frac{1}{d_L} - \frac{1}{d_S} \Bigg)},
    \label{eq:einstein radius}
\end{equation}
where ${d_S}$ is the observer-source distance, and ${d_L}$ is the observer-lens distance.

Microlensing parallax, $\pi_E$, causes asymmetries in the photometric light curve and is caused by differences in the relative location of the observer with respect to the source\hyp{}lens axis, as can happen when Earth orbits the Sun during long duration events:

\begin{equation}
\label{eq:piE}
    \pi_E = \frac{\pi_{rel}}{\theta_E},
\end{equation}
where $\pi_{rel}$ is the relative parallax:

\begin{equation*}
    \pi_{rel} = 1 \mathrm{AU} \bigg( \frac{1}{d_L} - \frac{1}{d_S} \bigg).
\end{equation*}

The Einstein crossing time can then be determined using the magnitude of the relative source\hyp{}lens proper motion, $\mu_{rel}$:

\begin{equation*}
    t_{E} = \frac{\theta_E}{\mu_{rel}},
    \label{eq:einstein crossing time}
\end{equation*}
which describes the characteristic timescale of the event (i.e., the time it takes for the source to transit the Einstein radius of the lens). 

In addition to photometric measurements, one can obtain astrometric measurements for microlensing events using the location and movement of stars. In the absence of lensing and parallax a source star would appear to make a straight line on the sky over most observable timescales, however, in the presence of lensing, the source appears to shift non\hyp{}linearly throughout the duration of the microlensing event. Thus, by measuring the change in location of the source image centroid throughout the microlensing event, one can determine the astrometric shift, $\vec{\delta_{c}}$:

\begin{equation}
\label{eq:astro shift}
    \vec{\delta_{c}} = \frac{\theta_{E}}{u^2+2} \vec{u}.
\end{equation}

Then, using the angular on\hyp{}sky positions of the source and lens ($\varphi_{S}$ and $\varphi_{L}$, respectively), one can determine the source-lens separation vector: 

\begin{equation*}
    \vec{u}=\frac{\vec{\varphi_{S}}-\vec{\varphi_{L}}}{\theta_{E}}.
\end{equation*}

By comparing the observed distribution of microlensing events to simulated distributions in $\pi_{E}$, $t_{E}$, $\bar{\delta}_{c}$, and $\mu_{rel}$ space, both with and without PBHs, we can ultimately place constraints on PBH dark matter abundance in the Milky Way. We can then estimate expected statistical signatures for each population by simulating the stellar and dark matter components of the Milky Way, and can analyze resulting microlensing properties after being filtered by a simulated observational survey.

In this work, we describe how \texttt{PopSyCLE} is used to generate a simulated microlensing survey in the Milky Way (\S \ref{sec:PopSyCLE Pipeline}), the steps used to inject PBHs into \texttt{PopSyCLE} (\S \ref{sec:Primordial Black Hole Injection}), and analysis on the resulting microlensing events (\S \ref{sec:microlensing sims}). As a proof\hyp{}of\hyp{}concept, we then consider our simulation results in the context of current and future surveys (\S\ref{sec:Results}).

\section{PopSyCLE Pipeline}
\label{sec:PopSyCLE Pipeline}

Population Synthesis Code for Compact Object Microlensing Events, or \texttt{PopSyCLE}\footnote{https://github.com/jluastro/PopSyCLE}, is a Python package that enables the user to simulate synthetic microlensing surveys in the Milky Way. \texttt{PopSyCLE} is a resolved microlensing simulation code, and the first of its kind to include both photometric and astrometric microlensing signals, and numerically derived compact object models \citep[][ henceforth Lam\citeyear{Lam:2020}]{Lam:2020}.

\texttt{PopSyCLE} relies on first using \texttt{Galaxia}\footnote{http://galaxia.sourceforge.net} \citep{Sharma:2011} to generate a synthetic model of the Milky Way populated with stars, and \texttt{SPISEA}\footnote{https://github.com/astropy/SPISEA } \citep{Hosek:2020} to generate and inject compact objects into the Milky Way model. \texttt{PopSyCLE} then uses the resulting outputs to simulate a microlensing survey with user\hyp{}specified parameters, enabling the user to control survey length, observing frequency, and various observational parameters. We briefly summarize these tools in the remainder of this section (additional detail in Lam\citeyear{Lam:2020}). 

\subsection{Galaxia}
\texttt{Galaxia} is first used to populate a stellar model of the Milky Way, using the Besan\c con analytic model \citep{Robin:2004}. We adopt changes to the bulge kinematics (following the ``v3" modifications from Lam\citeyear{Lam:2020} Appendix A) and reddening laws (Lam\citeyear{Lam:2020} Appendix B) in the standard \texttt{Galaxia} package in an attempt to eliminate discrepancies with observations, and to more accurately represent the event rate predictions for the scope of this study.

For bulge kinematics, we modify the Milky Way parameters in \texttt{Galaxia} to use a pattern speed of $\Omega = 40$ $km$ $s^{-1}$ $kpc^{-1}$, bulge velocity dispersion of $\sigma_{R} = \sigma_{\phi} = 100$ $km$ $s^{-1}$, bar angle of $\alpha = \ang{28}$, and bar length of $x_{0} = 0.7$ $kpc$. The default \texttt{Galaxia} extinction maps are not accurate in the high extinction regions of the inner bulge, so we modify our output in accordance to the reddening law from \citet{Damineli:16}.

\subsection{SPISEA}
The Stellar Population Interface for Stellar Evolution and Atmospheres, \texttt{SPISEA}, is a Python package used to generate star clusters (single\hyp{}age and single\hyp{}metallicity populations) using various input parameters, including age, mass, metallicity, extinction, atmospheric models, and initial mass functions \citep{Hosek:2020}. \texttt{SPISEA} allows for compact object generation by supporting user control of the initial\hyp{}final mass relation (IFMR) \citep{Rose:2022}. We follow the \texttt{SPISEA} modifications outlined in Lam\citeyear{Lam:2020} \S 2.2, which uses the \texttt{Raithel18} \citep{Raithel:2018} IFMR from recent simulations, along with the zero\hyp{}age main sequence (ZAMS) mass of each star, in order to build a population of stellar evolved black holes (BH), neutron stars (NS), and white dwarfs (WD). 

\begin{deluxetable*}{clc}[t]
\vspace{13pt}
\tabletypesize{\small}
\tablecaption{{PBH Simulation Parameters}
    \label{tab:PBH Parameters}}
\tablehead{
    \colhead{$\textbf{Parameter}$} & 
    \colhead{\hspace{0.2in}$\textbf{Description}$} &
    \colhead{\hspace{0.4in}$\textbf{Value}$}
}
\startdata
     $\bar{m}_{\text{PBH}}$ & \hspace{0.4in}Mean PBH mass & \hspace{0.4in}30 $M_{\odot}$ \\
     $f_{\text{DM}}$ & \hspace{0.4in}PBH Dark matter fraction & \hspace{0.4in}1.0 \\
     $v_{\text{esc}}$ &  \hspace{0.4in}Milky Way escape velocity & \hspace{0.4in}550 $km$ $s^{-1}$ \\
     $\rho_{0}$ & \hspace{0.4in}Characteristic density of Milky Way dark matter halo & \hspace{0.4in}0.0093 \(M_\odot\) $pc^{-3}$ \\
     $r_{s}$ & \hspace{0.4in}Scale radius of Milky Way dark matter halo & \hspace{0.4in}18.6 $kpc$ \\
     $r_{gc}$ & \hspace{0.4in}Distance between the Sun and Galactic center & \hspace{0.4in}8.3 $kpc$ \\
     $r_{\text{max}}$ & \hspace{0.4in}Max distance from Earth to populate PBHs ($2 \times r_{gc}$)& \hspace{0.4in}16.6 $kpc$ \\
     $\gamma$ & \hspace{0.4in}Milky Way Halo inner slope & \hspace{0.4in}1, 0.5, or 0.25 \\
\enddata
\end{deluxetable*}

\subsection{PopSyCLE}
\label{sec:PopSyCLE}

\texttt{Galaxia} assigns objects to four populations: thin disk ($150$ $Myr$ to $10$ $Gyr$), bulge ($10$ $Gyr$), thick disk ($11$ $Gyr$), and stellar halo ($14$ $Gyr$). \texttt{PopSyCLE} then splits the thin disk population into 29 age bins and changes the age of the stellar halo population to $13.8$ $Gyr$ (Lam\citeyear{Lam:2020} \S 3).

Once \texttt{SPISEA} has generated the compact objects\footnote{We use \texttt{PopSyCLE} at commit \texttt{\#018af97} which considers only singular stars and compact objects.}, they get injected back into the stellar population\footnote{\texttt{PopSyCLE} is being updated to support binary populations, updated IFMRs, and a NS mass spectrum, however, these updates are not reflected in the results of this work.}. Following Lam\citeyear{Lam:2020}, we assign initial ``kick" velocities (in addition to stellar velocities) to NSs and BHs, assigning values of 350 $km$ $s^{-1}$ and 100 $km$ $s^{-1}$, respectively. Compact object (BHs, WDs, NSs) position and velocity follow that of the stellar population in \texttt{Galaxia}, and are considered ``dark" sources (with the exception of luminous white dwarfs) (Lam\citeyear{Lam:2020}). 

Microlensing events are assumed to be ``point source, point lens" (meaning both the source and lens are treated as point sources), and event calculation follows that described in Lam\citeyear{Lam:2020} \S{4}. Initial event candidates are determined using a cut on the separation between each object and its nearest on\hyp{}sky neighbor, $\Delta\theta \leq \theta_{blend}$, in which $\theta_{blend}$ is the survey specific seeing\hyp{}disk radius, and defines the angular projected distance from the lens in which objects are considered blended. 

Event detectability is determined by calculating the Einstein radius, $\theta_{E}$, following Equation \ref{eq:einstein radius}. The list of candidates is then narrowed down to contain only those events which satisfy the following equation:

\begin{equation*}
    \frac{\Delta\theta}{\theta_E} < u_{0,max},
\end{equation*}
where $u_{0,max}$ is the impact parameter threshold for the survey. Events are then cut again to ensure that they \change{satisfy the user-specified selection criteria, related to the survey duration and cadence}. Lastly, events with dark objects as sources are removed, as they are non\hyp{}observable events. Every event candidate remaining is considered an event in the simulation and the user is left with catalogs containing photometric and astrometric microlensing parameters for each event.

\section{Primordial Black Hole Injection}
\label{sec:Primordial Black Hole Injection}

Before this work, \texttt{PopSyCLE} supported the compact object populations of NSs, WDs, and BHs, but had no implementation for PBHs. We update \texttt{PopSyCLE} to include the addition of an optional PBH population\footnote{https://github.com/jluastro/PopSyCLE/tree/pbh} using the parameters and assumptions given in Table \ref{tab:PBH Parameters}.

Including a PBH population in \texttt{PopSyCLE} requires injecting PBHs into our projected survey line\hyp{}of\hyp{}sight (LOS) by first determining how many PBHs to simulate for the given field (detailed in \S \ref{sec:PBH Number}). PBHs have spatial and velocity distributions following that of the dark matter halo, rather than the distributions for stellar remnant compact objects which were born in the galactic disk. For this reason, we then have to assign positions (detailed in \S \ref{sec:PBH Positions}) and velocities (detailed in \S \ref{sec:PBH Velocities}) to each PBH based on the halo density and speed profiles. Three\hyp{}dimensional positions and velocities are assigned by sampling the radial components (with respect to the Galactic center) from the relevant distributions, and randomly sampling the angular components over the possible ranges. Once a PBH population for a given field has been generated, we inject the objects into the synthetic Milky Way generated from \texttt{Galaxia} and \texttt{SPISEA}, thus enabling \texttt{PopSyCLE} to simulate a microlensing survey with PBHs present.

\subsection{PBH Distribution}
\label{sec:DM Dist}

\begin{figure}[t]
    \begin{center}
        \includegraphics[width=.49\textwidth]{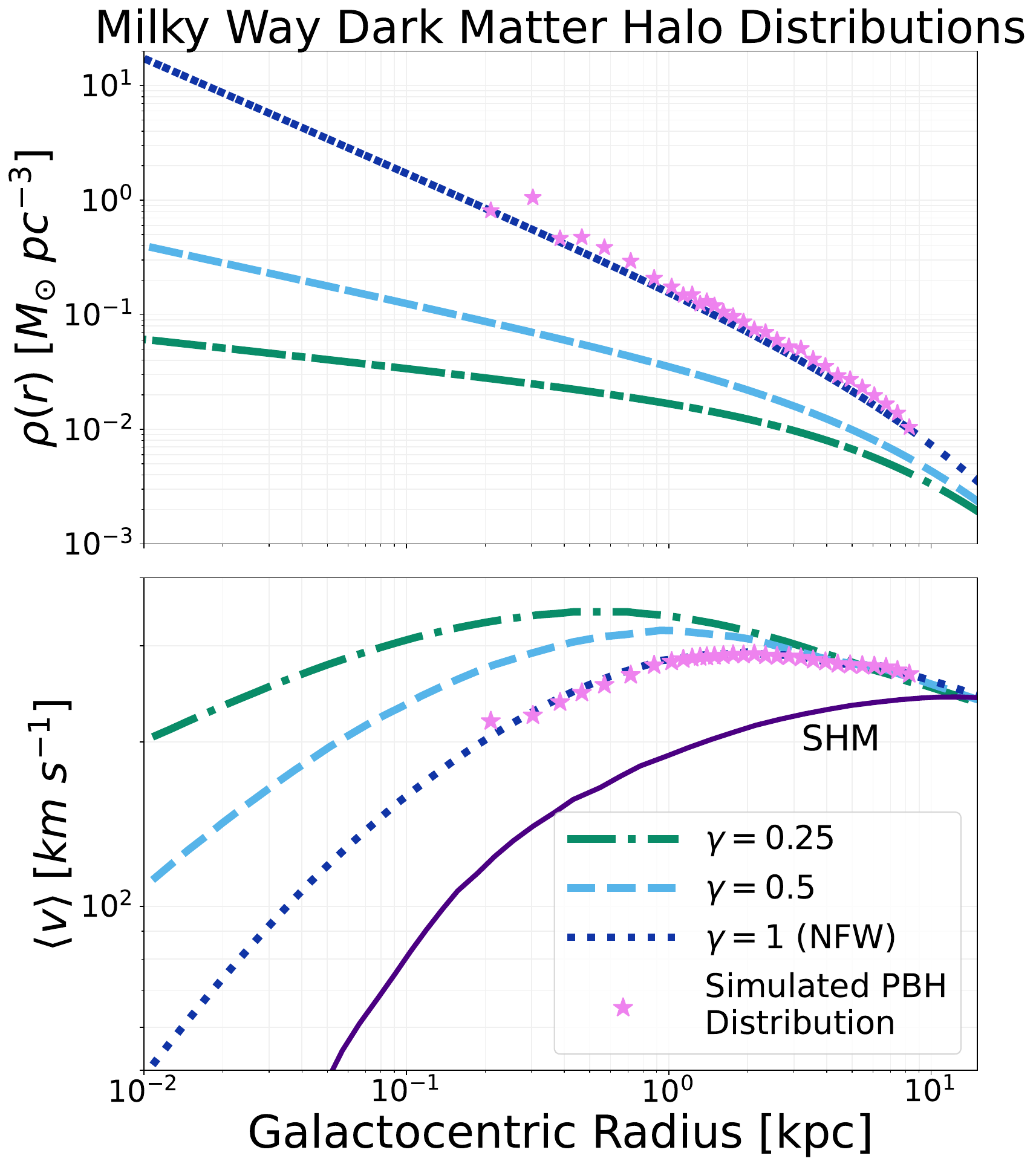}
        \caption{\textit{Upper}: Halo density as a function of radius, for varying halo inner density slopes. \textit{Lower}: A recreation of the Eddington inversion curves from Figure 11 of \citet{Lacroix:2018}, demonstrating how the mean speed profile changes as a function of inner slope. The standard halo model is shown by the solid line. In both panels the stars show that the injected PBHs trace the intended profiles.}
        \label{fig:gamma}
    \end{center}
\end{figure}

We assume that Milky Way dark matter follows the halo density profile in \citet{McMillan:2016}:

\begin{equation}
    \label{eq:rho}
    \rho \left(r\right) = \frac{\rho_{0}}{\left(\frac{r}{r_s}\right)^{\gamma}\left(1 + \frac{r}{r_s}\right)^{\left(3-\gamma\right)}},
\end{equation}
where $r$ is the distance from the galactic center (in the galactocentric coordinate frame), $\gamma$ is the halo inner density slope, $r_s$ is the scale radius of the halo, and $\rho_0$ is the halo characteristic density. An inner slope of $\gamma=1$ corresponds to the Navarro\hyp{}Frenk\hyp{}White (NFW) profile \citep{Navarro:1996}, and is used for all simulation results in this work (though \texttt{PopSyCLE + PBH} supports $\gamma$ values of 1, 0.5, and 0.25). The upper panel of Figure \ref{fig:gamma} demonstrates how the halo density changes as a function of radius and $\gamma$. We choose $r_{gc}=8.3$ $kpc$ to be the distance between the Sun and Galactic center \citep[which is typical of the literature, e.g.,][]{Gillessen:2009, Chatzopoulos:2015}, $r_s = 18.6$ $kpc$ to be the halo scale radius, $\rho_{0} = 0.0093$ $M_{\odot}$ $pc^{-3}$ to be the halo characteristic density \citep[which are the median $r_s$ and $\rho_{0}$ values determined by][given their mass model]{McMillan:2016}, and $r_{\text{max}} \equiv 2 \; r_{gc}$ to be the maximum galactocentric radius (towards the galactic center) to populate objects out to (as we do not expect to detect a significant number of microlensing events beyond that).

\subsection{PBH Number}
\label{sec:PBH Number}

\begin{figure*}[t]
    \begin{center}
        \includegraphics[width=.99\textwidth]{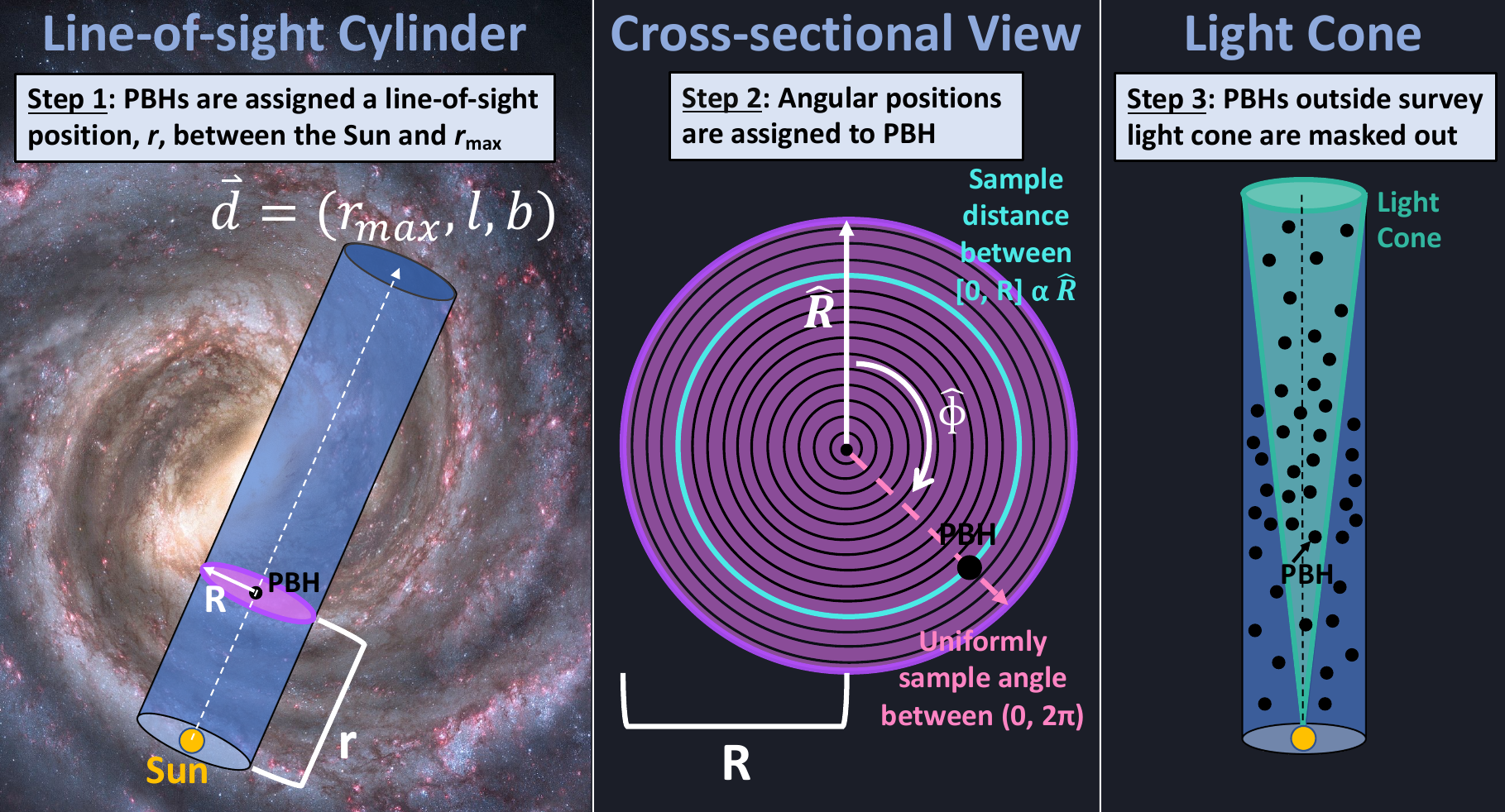}
        \caption{PBH position assignment for a single PBH in a simulation \emph{Left}: A cylinder is constructed with radius $R$ (the survey radius projected at $r_{max}$), and centered on the LOS (dotted line \change{extending from the Sun to the field center}). PBHs are assigned a distance, $r$, between [0, $r_{max}$], by uniformly sampling the cumulative distribution function of Equation \ref{eq:rho} \change{using standard inverse transform sampling methods}. \emph{Middle}: A cross\hyp{}sectional view of the \change{LOS cylinder for the location assignment of a single PBH}. Three\hyp{}dimensional positions within the \change{LOS} cylinder are assigned by randomly sampling a radial distance \change{$\hat{R}$} between [0, $R$] \change{(proportional to the cross-sectional area, $\pi \hat{R}^2$)}, and \change{uniformly sampling} an angle between [0, 2$\pi$]. This is repeated until each PBH has a 3D position within the cylinder. \emph{Right}: PBHs that fall outside of the survey light cone are masked out.}
        \label{fig:visual}
    \end{center}
\end{figure*}

\change{To determine the number of PBHs to include in a given simulation, we first construct a LOS extending from $r=0$ (the Sun) to the center of the field being simulated ($l$ and $b$, at a heliocentric distance of $r = r_{\text{max}}$). We then assume that PBHs will populate a heliocentric LOS cylinder (shown in Figure \ref{fig:visual}, described in \S \ref{sec:PBH Positions}) whose center axis is the LOS, and radius, $R$, corresponds to the projected survey radius at the distance $r_{\text{max}}$ (see Figure \ref{fig:visual}). Next, we split the LOS into 100,000 equally spaced points for numerically integrating the LOS dark matter density (accounting for galactocentric to LOS coordinate transformations), enabling us to approximate the total mass within the LOS cylinder, $M_{\text{LOS}}$\footnote{This approximation becomes less accurate as $R$ increases, however, we expect this error to be sub\hyp{}dominant to other approximations, as we simulate at most 0.34 $deg^2$.}.}

Using the fraction of dark matter composed of PBHs, $f_{\text{DM}}$, and mean PBH mass, $\bar{m}_{\text{PBH}}$, we can then determine the number of PBHs to inject in a given field, $N_{\text{PBH}}$:

\begin{equation}
    \label{eq:num pbh}
    N_{\text{PBH}} = f_{\text{DM}} \left(\frac{M_{\text{LOS}}}{\bar{m}_{\text{PBH}}} \right)
\end{equation}

For the scope of this work we assume $f_{\text{DM}}=1$ and assign PBHs a monochromatic mass of $\bar{m}_{\text{PBH}}=30$ $M_{\odot}$, as we find this to be roughly consistent with inferred masses of LIGO detected black holes \citep[e.g.,][]{Abbott:2016, Abbott:2017a, Abbott:2017b, Lu:2019}. We acknowledge that these assumptions contradict existing $f_{\text{DM}}$ constraints  \citep[e.g.,][]{Blaineau:2022, 2001:Alcock, Tisserand:2007, Wyrzykowski:2011}, and expectations of an extended mass distribution \citep[e.g.,][]{Chapline:1975a, Chapline:1975b, Chapline:2016, Chapline:2018, Carr:1975, Carr:2018, Carr:2021}, however, our primary focus is on demonstrating the addition of a PBH population into the \texttt{PopSyCLE} simulation code. For this reason, we believe these choices are acceptable, as they enable readers to easily transform the results to other values of $f_{\text{DM}}$ and $\bar{m}_{\text{PBH}}$, while reducing Poisson noise.

\subsection{PBH Positions}
\label{sec:PBH Positions}

\change{After determining the number of PBHs that should be simulated within the LOS for a given field, $N_{\text{PBH}}$, we assign each PBH a distance along the LOS, $r$, using inverse transform sampling \citep[e.g.,][]{Press:1988} to assign each PBH a galactocentric radius based on a numerical approximation of the cumulative distribution function of the normalized dark matter density function. Each PBH galactocentric radius is then converted to a heliocentric distance along the LOS, $r$, which falls directly on the LOS (see left panel of Figure \ref{fig:visual} for a single PBH example).}


\change{Each PBH then needs two additional dimensional quantities for a final 3D position within the LOS cylinder. We project each PBH onto a circular cross-section of the LOS cylinder at sampled distance $r$, then, assuming an isotropic projection distribution, we sample radial and angular component values for each PBH (see middle panel of Figure \ref{fig:visual} for a single PBH example). For each PBH, radial distance (from the center axis of the LOS cylinder to the projected field radius, $R$) is sampled proportional to the cylindrical cross-sectional area ($\pi \hat{R}^2$), while an angle is uniformly sampled between [0, 2$\pi$]\footnote{This cumulative integration technique neglects correlation between mass and dark matter halo location (as dynamical friction forces on an extended PBH mass spectrum might introduce).}.} 

\change{The result of this sampling process is a cylinder along the LOS populated with the number of PBHs determined by Equation \ref{eq:num pbh}, following that of the halo profile in Equation \ref{eq:rho}. We performed convergence checks on the discritized space and find that as you pass over the galactic center, distances within $\approx 0.157$ $kpc$ of the center cause the density to converge, so for our simulations we do not go below this distance radially from the galactic center. Before velocity assignment, we mask out PBHs that cannot be observed by cutting out PBHs which fall outside of the survey light cone (i.e., solid angle). An example of this cut is shown in the right panel of Figure \ref{fig:visual}.}

\subsection{PBH Velocities}
\label{sec:PBH Velocities}

Each PBH is assigned a mean speed $\langle \text{v} \rangle$ given its galactocentric radius, using the Eddington formalism \citep{Eddington:1916} for the halo velocity as derived in \citet{Lacroix:2018}. The Eddington inversion model (which relates total gravitational potential and mass density to the dark matter phase\hyp{}space distribution) deviates from the standard halo model (SHM; in which velocities follow a truncated Gaussian), especially at smaller Galactic radii (see ``SHM" curve on lower panel of Figure \ref{fig:gamma}). However, \citet{Lacroix:2018} think the Eddington model more representative of PBH velocities, thus, we use the NFW parameterization of this relationship (illustrated by the $\gamma=1$ panel on Figure \ref{fig:gamma}).

Next, we assume a Maxwellian velocity distribution, and for each PBH we use its assigned $\langle \text{v} \rangle$ as the distribution mean, to get the Maxwell distribution parameter: 

\begin{figure}[p!]
    \includegraphics[width=.47\textwidth]{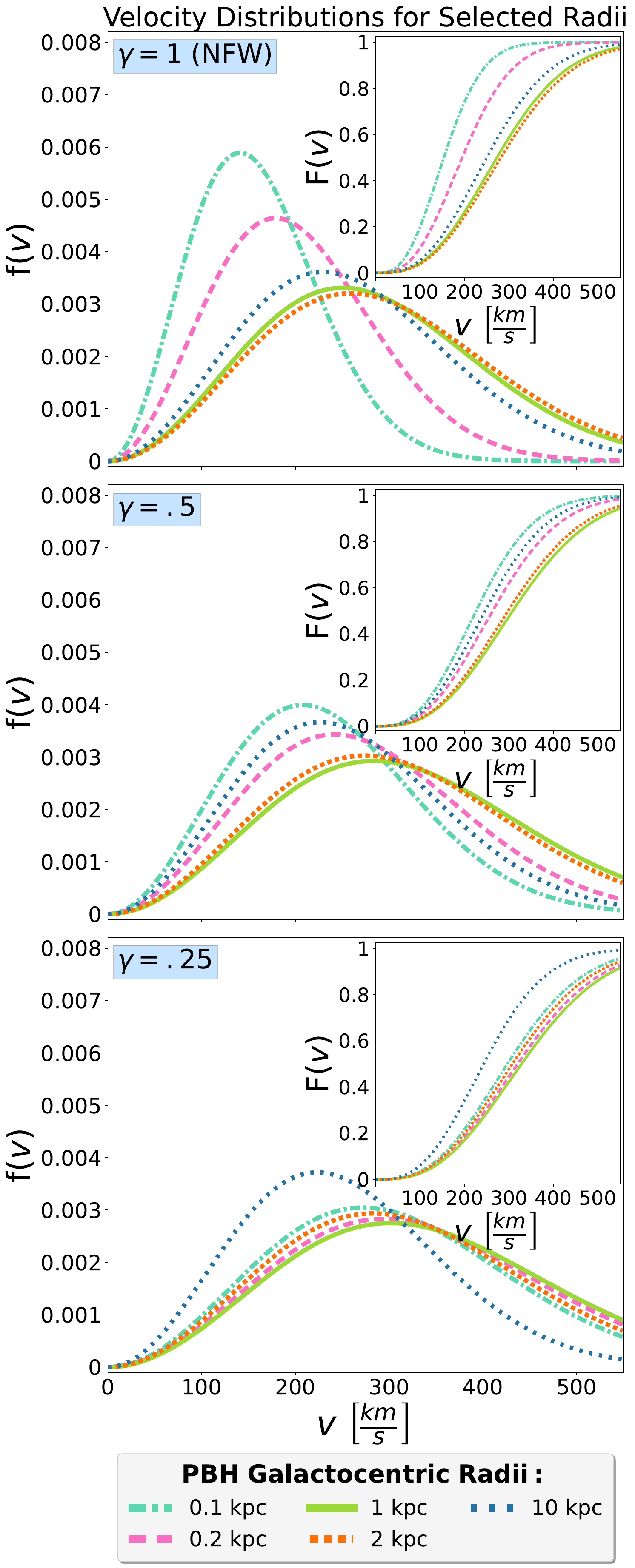} 
    \caption{\change{Velocity magnitude distributions for selected PBH radii, at three different $\gamma$ values.}}
    \label{fig:velocity distribution}
\end{figure}

\begin{equation*}
    a = \frac{1}{2}  \langle \text{v} \rangle \sqrt{\frac{\pi}{2}},
\end{equation*}
then calculate the probability density function:

\begin{equation}
    \label{eq:f}
    f(v) = \sqrt{\frac{2}{\pi}} \frac{v^{2} e^{-\frac{v^{2}}{2a^{2}}}}{a^{3}},
\end{equation}
in which $v$ is the root\hyp{}mean square (RMS) velocity. We uniformly sample a single value from the cumulative distribution function \change{(again, using inverse transform sampling)}: 

\begin{equation}
    \label{eq:F}
    F(v) = \int f(v) = \text{erf} \left( \frac{v}{a \sqrt{2}} \right) - \sqrt{\frac{2}{\pi}} \frac{v e^{\frac{-v^{2}}{2 a^2}}}{a},
\end{equation}
where \emph{erf} is the error function. Each PBH is assigned the corresponding sampled RMS velocity.

We let $v$ take values between 0 and the escape velocity $v_{esc}$ = 550 $km$ $s^{-1}$, which is consistent with current estimates, such as those from the Radial Velocity Experiment (RAVE) survey \citep{Smith:2007, Piffl:2014}. For PBHs with $\langle \text{v} \rangle < 250$  $km$ $s^{-1}$, this $v_{esc}$ cut filters out $<$ 1\% of the PBH distribution, while PBHs with mean speeds closer to the maximum, $\langle \text{v} \rangle$ $\sim{300}$ $km$ $s^{-1}$, equates to a loss of $\sim$3.5\%. Because the number of PBHs in the LOS cylinder was previously selected to match the total mass of the cylinder, the number of PBH lensing events will be biased slightly low. Figure \ref{fig:velocity distribution} illustrates $f(v)$ and $F(v)$ distribution examples (in the main figures and subfigures, respectively) for a few selected galactocentric PBH radii and $\gamma$ values.

Following similar projection methods as described in \S \ref{sec:PBH Positions}, we take our velocity magnitude $v$, and randomly sample two angular components for latitudinal and longitudinal velocity components\update{\footnote{We assume a Maxwellian distribution for PBH velocities as a computational shortcut, but use mean speeds following the Eddington inversion results from \citet{Lacroix:2018} (which assumes velocities deviate from the truncated Maxwellian of the SHM). However, we do not expect this simplifying assumption to be a dominant source of uncertainty in our analysis.}}. \change{The longitudinal component is sampled uniformly between [0, $2\pi$], while the latitudinal component is sampled uniformly on the unit sphere (i.e., uniform in $\cos({\rm latitude})$ between [$-1$, $1$]).}

\subsection{Other PBH Parameters}

In addition to six\hyp{}dimensional kinematics, \texttt{PopSyCLE} requires additional parameters for microlensing survey simulation. \change{Because PBHs form in the early universe and not through astrophysical formation channels, they do not have a ZAMS mass. However, to ensure smooth integration into \texttt{PopSyCLE}, the PBH mass at the time of the simulation is assigned as the ZAMS mass.} We add a population type for PBHs dubbed ``dark matter halo" (in addition to those supplied by \texttt{Galaxia}) and assign an age of 13.8 Gyr. In addition to the \texttt{PopSyCLE} remnant IDs of 0 (star), 101 (WD), 102 (NS), and 103 (BH), a new remnant ID of 104 is assigned to PBHs. Magnitudes and photometric information does not need to be considered, as PBHs are ``dark". 

\section{Microlensing Simulations}
\label{sec:microlensing sims}

Each of our mock surveys closely resemble observational criteria for the survey being simulated. The \texttt{PopSyCLE} survey parameters used are listed in Table \ref{tab:Survey Parameters} and include: duration (length of the survey), cadence (time between each observation), $\theta_{blend}$ (seeing\hyp{}disk radius) and $u_{0, \text{max}}$ (maximum impact parameter) as described in \S \ref{sec:PopSyCLE}, area (surface area of the survey field\hyp{}of\hyp{}view projected on the sky), limiting magnitude of the survey, optical filter for the survey, $\Delta_{mag}$ (``bump magnitude"; the difference between baseline magnitude and peak magnitude, used to remove low\hyp{}amplitude events), and $t_e$ (the Einstein crossing time described in \S\ref{sec:Intro}). For this work we assume an NFW dark matter profile, and use $\gamma=1$ when generating the PBH population. \update{\texttt{PopSyCLE} and \texttt{SPISEA} use the Vega magnitude system (denoted by subscript \emph{V}), which we convert to AB magnitudes (denoted by subscript \emph{AB}) where appropriate using values from Table 1 of \citet{Blanton:2007}.}

\texttt{PopSyCLE} cadence and actual observational cadence do not exactly correlate, as \texttt{PopSyCLE} spreads out observations over \change{the user-specified duration, and with the user-specified cadence, not necessarily} accounting for daytime or seasonal gaps. Our OGLE cadences are derived from the reported observations per field each night, and Roman cadences are determined by spreading the expected microlensing survey over the duration of telescope operation, meaning we neglect these daytime and seasonal gaps as well (discussed further in \S \ref{sec:ogle iv sim} and \S \ref{sec:roman sims}). This choice of cadence may lead to underestimating the number of events with timescales shorter than the specified cadence, but we find this affect negligible. 

\begin{deluxetable}{|l|c|c|c|}[t]
\tabletypesize{\small}
\tablecaption{{\texttt{PopSyCLE} Survey Parameters}
    \label{tab:Survey Parameters}}
\tablehead{
    \multicolumn{1}{|l|}{{}} & 
    \multicolumn{2}{c|}{\textbf{Mock OGLE\hyp{}IV}} &
    \multicolumn{1}{c|}{\textbf{Mock}} \\
    \cline{2-3}
    \multicolumn{1}{|l|}{\textbf{Parameter}} & 
    \multicolumn{1}{c|}{\textbf{EWS Cut}} &
    \multicolumn{1}{c|}{\textbf{Mr\'{o}z19 Cut}} &
    \multicolumn{1}{c|}{\textbf{Roman}}
}
\startdata
     Duration & 2920 days & 2920 days & 1825 days \\
     Cadence$^{*}$ & $\sim$2.5 days$^{\dagger}$ & $\sim$2.5 days$^{\dagger}$ & $\sim$3 days \\
     $\theta_{blend}$ & 0.65" & 0.65" & 0.09" \\ 
     $u_{0, \text{max}}$ & $\leq$ 2 & $\leq$ 1 & $\leq$ 2 \\
     Area & 0.34 $\text{deg}^2$ & 0.34 $\text{deg}^2$ & 0.16 $\text{deg}^2$ \\
     Min. Mag.$^{\ddagger}$ & $\leq$ \update{21.0} & $\leq$ \update{21.0} & $\leq$ \maybe{26.7} \\
     Filter & $I_{\color{black}{V}}$ & $I_{\color{black}{V}}$ & $H_{\color{black}{AB}}$ \\
     $\Delta_{mag}$ & $\geq$ 0.1 & -- & $\geq$ 0.1\\
     $t_E$ & -- & $0.5 \leq t_E \leq 300$ & --
\enddata
\tablecomments{$^{*}$\texttt{PopSyCLE} defined cadence, which does not account for daytime gaps or survey gaps.\\
$^{\dagger}$Fields BLG\hyp{}500 and BLG\hyp{}611 use a \texttt{PopSyCLE} survey cadence of 10 days.\\
$^{\ddagger}$Mock EWS and Mock Roman use minimum baseline magnitude. Mock Mr\'{o}z19 uses minimum source magnitude.}
\end{deluxetable}

\begin{figure*}[p]
    \begin{center}
        \includegraphics[width=.99\textwidth]{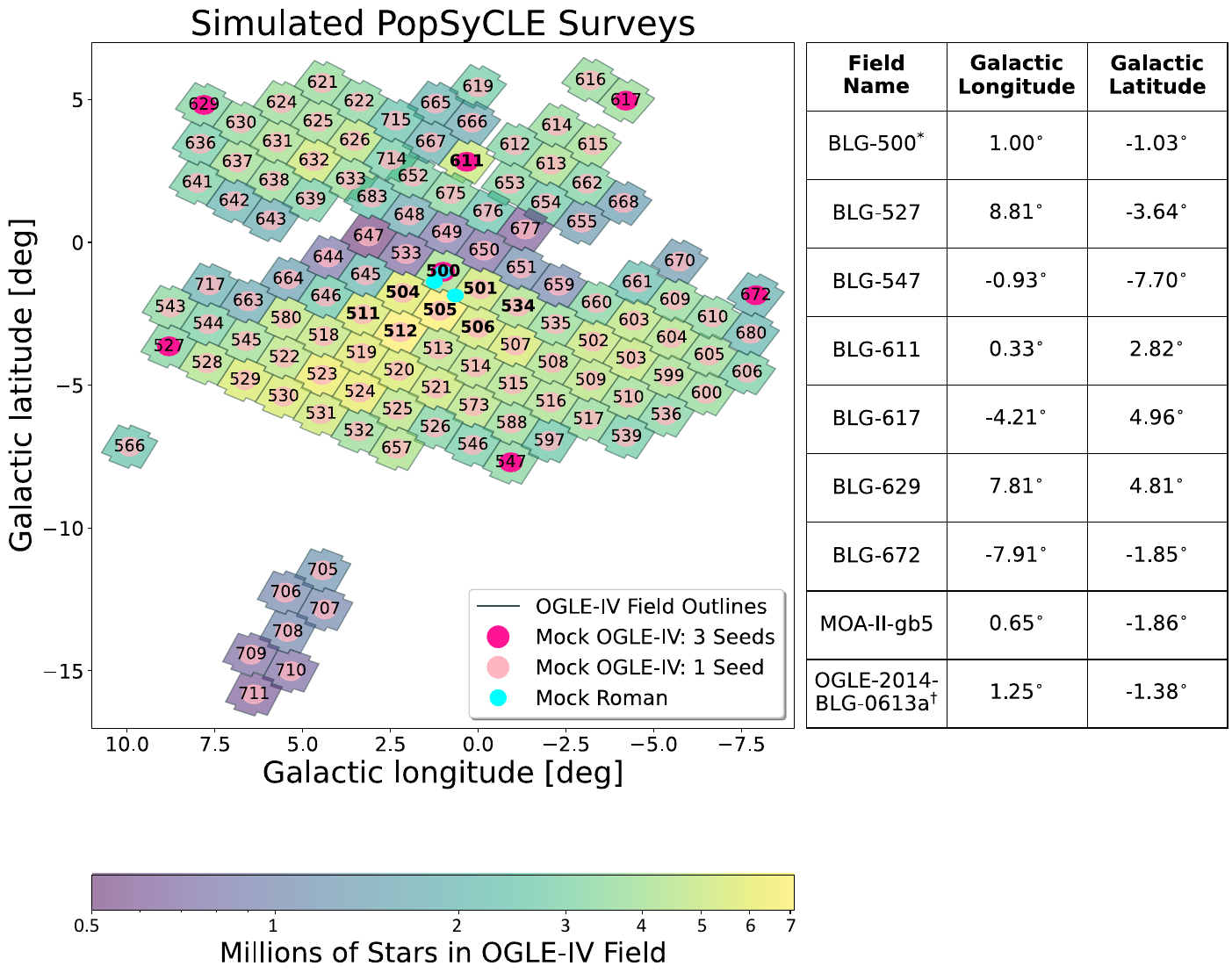}
        \caption{\emph{Left:} Simulated survey areas (drawn to scale) plotted over the OGLE\hyp{}IV survey footprint. The background color scale represents the number of stars in a given field, and the number on each field is the OGLE bulge (BLG) field number. \change{Each light pink field was simulated with one seed, while the dark pink fields were simulated with three seeds each. The fields that have their numbers in bold are the high cadence OGLE-IV fields (as defined in \citet{Mroz:2019}), and everything else is considered a low cadence field.} \emph{Right:} The name and Galactic coordinates for each \change{field that was simulated three times (}each field is centered on the given coordinates, and has a simulated area specified in Table \ref{tab:Survey Parameters}).\\ $^{*}$BLG\hyp{}500 is both a 0.34 $\text{deg}^2$ Mock OGLE\hyp{}IV field and 0.16 $\text{deg}^2$ Mock Roman field.\\
        $^{\dagger}$OGLE\hyp{}2014\hyp{}BLG\hyp{}0613a is centered on a microlensing event \change{reported by} OGLE EWS.}
        \label{fig:PopSyCLE fields}
    \end{center}
\end{figure*}

\change{For each survey, we specify a set of fields to run \texttt{PopSyCLE} on. Fields and locations (with the area from Table \ref{tab:Survey Parameters} drawn to scale) are detailed in Figure \ref{fig:PopSyCLE fields}. Each field in Figure \ref{fig:PopSyCLE fields} is run one time at minimum, where light pink denotes each simulated OGLE-IV field that is run only once, dark pink denotes each OGLE-IV field that is run over three random seeds each, and cyan denotes each simulated Roman field that is run over three random seeds each. The dark pink and cyan fields are 9 fields taken from Lam\citeyear{Lam:2020}, in which each field is centered at a Galactic longitude and Galactic latitude ($l$ and $b$, respectively, noted on the right panel of Figure \ref{fig:PopSyCLE fields}) that either corresponds to the center of a bulge field from the Optical Gravitational Lensing Experiment (OGLE) survey \citep{Udalski:1992} or the Microlensing Observations in Astrophysics (MOA) survey \citep{Muraki:1999}. Only a 0.34 $\text{deg}^2$ or 0.16 $\text{deg}^2$ (for OGLE-IV and Roman, respectively) portion of each survey footprints is simulated, then extrapolation is done (when necessary) with the assumption that stellar density is constant over each field. Fields with a label in bold on Figure \ref{fig:PopSyCLE fields} are ``high-cadence" OGLE-IV fields (as defined in \citet{Mroz:2019}), while everything else is considered a low cadence field.}

\subsection{Mock OGLE\hyp{}IV Simulations}
\label{sec:ogle iv sim}

The OGLE survey \citep{Udalski:1992} is designed to detect microlensing events in the Milky Way's bulge and Magellanic Clouds by frequently surveying the sky to look for microlensing signatures. OGLE phase 1 (OGLE\hyp{}I) began in 1992, and recently finished the fourth phase (OGLE\hyp{}IV), making observations with the 1.3 $m$ Warsaw telescope at the Las Campanas Observatory in Chile over an 8 year period \citep{Udalski:2015}.

Our ``Mock OGLE\hyp{}IV" surveys use the corresponding values from Table \ref{tab:Survey Parameters} to simulate the portions of the OGLE microlensing survey shown in \change{light pink and dark pink} on Figure \ref{fig:PopSyCLE fields}. We adopt these field parameters from the ``Mock EWS\footnote{EWS = Early Warning System. The EWS reduces OGLE data in real time to alert on potential microlensing events. We base parameter estimates on reported numbers from 2016\hyp{}2018.}" and ``Mock Mr\'{o}z19" surveys from Lam\citeyear{Lam:2020}. The Mr\'{o}z19 cut returns OGLE\hyp{}IV events that might be detectable, and that are also of sufficient signal\hyp{}to\hyp{}noise such that their lensing parameters could be reasonably estimated. The EWS cut includes additional events that could conceivably be detected with OGLE, but that one may not be able to get reliable lensing parameter estimates for.

Using Table 6 from \citet{Mroz:2019} we calculate the average cadence for each OGLE field we simulate, then, averaging over all fields, we decide on a simulated survey cadence of 1 observation every $\sim$2.5 days (with the exception of BLG\hyp{}500 and BLG\hyp{}611 which use a cadence of 10 days due to computational constraints). Cadence affects the sensitivity of shorter scale events (in which $t_{E}$ is less than the time between observations) because the event will fall entirely between observations. While our simulated cadences are longer than some of the actual OGLE\hyp{}IV field cadences, Lam\citeyear{Lam:2020} show that a cadence of $\leq$ 10 days is sufficient to provide an accurate prediction of the $t_E$ distribution peak. 

All quantities reported from our \texttt{PopSyCLE} simulations are considered as corrected, as \texttt{PopSyCLE} does not account for events that fall between gaps, noisy observations, etc. 

To calculate event parameters with and without PBHs, we first determine the number of detectable stars in each field\change{, $N_s$,} by cutting the full object catalog from \texttt{Galaxia} on the limiting magnitude of the survey ($\color{black}I_V=21$). \change{To estimate the number of microlensing events (not including PBHs) in each field for each OGLE-IV cut, including relevant scaling and correcting for the actual survey efficiencies, we use the following equation:}

\begin{equation}
    \label{eq:ev Mroz}
    N_{\text{ev}} = N^{sim}_{\text{ev}} \left(\frac{1.4}{0.34}\right) \left(\frac{N^{M19}_{star}}{N^{M19}_{s}}\right) \langle \epsilon \rangle
\end{equation}
\change{where $N^{sim}_{\text{ev}}$ is the number of simulated microlensing events (excluding PBHs) output from \texttt{PopSyCLE} (after the cuts from Table \ref{tab:Survey Parameters} are applied), (1.4/0.34) scales our simulated number of microlensing events up to the full field area, $N^{M19}_{star}$ is the number of stars being monitored in that particular OGLE-IV field (values taken from Table 6 of \citet{Mroz:2019}), $N^{M19}_{s}$ is the number of sources in a field that are brighter than $\color{black}I_{V} = 21$ (with blends accounted for; values from Table 7 of \citet{Mroz:2019}), and $\langle \epsilon \rangle$ is the mean efficiency averaged over all of our simulated microlensing events in that field (i.e., all events in $N^{sim}_{\text{ev}}$). The quantity ($N^{M19}_{stars}$/$N^{M19}_{s}$) serves as a luminosity completeness correction factor}. 

\change{Then, to estimate the number of PBH microlensing events in each field, with relevant survey and efficiency corrections included, for each OGLE-IV cut, we use the following equation:}

\begin{equation}
    \label{eq:PBH Mroz}
    \begin{multlined}
        N_{\text{ev, PBH}} =\\
        N^{sim}_{\text{ev, PBH}} \left(\frac{N^{obs}_{\text{ev}}}{N^{sim}_{\text{ev}}\langle \epsilon \rangle+N^{sim}_{\text{ev, PBH}}\langle \epsilon' \rangle}\right) \langle \epsilon' \rangle
    \end{multlined}
\end{equation}
\change{where $N^{sim}_{\text{ev, PBH}}$ is the number of simulated PBH microlensing events output from \texttt{PopSyCLE} (after the cuts from Table \ref{tab:Survey Parameters} are applied), $N^{obs}_{\text{ev}}$ is the number of detected events listed in Table 7 of \citet{Mroz:2019} or the number of events reported between the 2011 and 2018 EWS seasons (depending on which cut we are determining $N_{\text{ev, PBH}}$ for), $N^{sim}_{\text{ev}}$ is the number of simulated microlensing events (excluding PBHs) output from \texttt{PopSyCLE} (after the cuts from Table \ref{tab:Survey Parameters} are applied), $\langle \epsilon \rangle$ is the mean efficiency averaged over all of our simulated microlensing events (excluding PBHs) in that field (i.e., all events in $N^{sim}_{\text{ev}}$), and $\langle \epsilon' \rangle$ is the mean efficiency averaged over each of our simulated PBH microlensing events in that field (i.e., all events in $N^{sim}_{\text{ev, PBH}}$). $N^{sim}_{\text{PBH}}$ and ($N^{sim}_{\text{ev}}+N^{sim}_{\text{ev, PBH}}$) both have a dependence on field area and corrections due to luminosity completeness, meaning that both the area and luminosity correction ratios cancel out and do not need to be considered for these calculations.}

\change{Efficiencies are calculated using each event $t_E$, with the published $t_E$ vs. efficiency binned values from the OGLE-IV survey\footnote{\change{https://www.astrouw.edu.pl/ogle/ogle4/\\microlensing\_maps/eff/}}. For simulated Mr\'{o}z19 fields that do not have PBH events, $\langle \epsilon' \rangle$ is taken to be the average efficiency value across all other fields that do have PBHs, $\langle \epsilon' \rangle = 0.268$. For simulated Mr\'{o}z19 fields that do not have any events (i.e., the number of NS, BH, WD, Star, and PBH events are all zero), $\langle \epsilon \rangle$ is taken to be the average efficiency value across all other fields that have non-zero events, $\langle \epsilon \rangle = 0.172$.}

\change{EWS efficiencies are unknown, therefore, we assume $\langle \epsilon \rangle = \langle \epsilon' \rangle$. However, when calculating the number of events (excluding PBHs) for our EWS simulations, we repeat the process of finding $\langle \epsilon \rangle$ using the reported OGLE-IV efficiency values, and fields that do not have any simulated events get assigned the average efficiency value across all other fields with non-zero events: 0.169.}

The event rate, $\Gamma$ (which excludes PBH events), is calculated by dividing the number of events without PBHs, $N_{\text{ev}}$, by \change{the number of stars that fall above the limiting magnitude for the given survey}, $N_s$, and then divide by the given survey duration. \change{We report our simulated OGLE\hyp{}IV results in Table \ref{tab:OGLE Comparison All}, which compares values from our analysis, with relevant quantities (i.e. $N_s$ and $N_{\text{ev}}$) scaled up from 0.34 $\text{deg}^2$ to the full field area, 1.4 $deg^2$, for equivalent comparison to their observational counterparts. We list $N_s$, $\Gamma$, $N_{\text{ev}}$, and $N_{\text{ev, PBH}}$ for our mock Mr\'{o}z19 and Mock EWS cuts, and list $N_s$, $\Gamma$, and $N_{\text{ev}}$ observational results reported by \citet{Mroz:2019} or the EWS website. $N_s$ and $\Gamma$ are completeness corrected values, while $N_{\text{ev}}$ and $N_{\text{ev, PBH}}$ are not, but are appropriately scaled by the various factors listed in Equations \ref{eq:ev Mroz} and \ref{eq:PBH Mroz} to make non-completeness corrected values to compare to their OGLE-IV observational counterparts. Three values are listed for fields that are run over three seeds each, and Poisson noise is assumed for all simulated values (this is because the exact noise profile is hard to characterize, and our sample of seeds follows Poisson uncertainty).}

\startlongtable
\begin{deluxetable*}{|c||c|c||c|c|c|c||c|c|c|c||c|c|}
\tabletypesize{\small}
\tablecaption{{\change{OGLE-IV Event Statistic Comparison}}
    \label{tab:OGLE Comparison All}}
\tablehead{
    \multicolumn{1}{|c||}{\textbf{OGLE}} & 
    \multicolumn{2}{c||}{$N_{s}$ [$10^6$]} &
    \multicolumn{4}{c||}{$\Gamma$ [$10^{-6}$ $\text{star}^{-1}$ $\text{year}^{-1}$]} &
    \multicolumn{4}{c||}{${N_{\text{ev}}}$} &
   \multicolumn{2}{c|}{${N_{\text{ev, PBH}}}$} \\
   \cline{2-13}
   \multicolumn{1}{|c||}{\textbf{Field}} &
    \multicolumn{1}{c|}{Mr\'{o}z19} &
    \multicolumn{1}{c||}{M/E} &
    \multicolumn{1}{c|}{Mr\'{o}z19} &
    \multicolumn{1}{c|}{M} &
    \multicolumn{1}{c|}{EWS} &
    \multicolumn{1}{c||}{E} &
    \multicolumn{1}{c|}{Mr\'{o}z19} &
    \multicolumn{1}{c|}{M} &
    \multicolumn{1}{c|}{EWS} &
    \multicolumn{1}{c||}{E} &
    \multicolumn{1}{c|}{M} &
    \multicolumn{1}{c|}{E}
}
\startdata
    BLG-500 & 6.78 & 4.71 & 23.9 & 23.5 & 110.9 & 134.42 & 164 & 111.52 & 752.0 & 646.23 & 7.01 & 20.83 \\
        &&&& 23.59 && 134.77 && 112.04 && 648.33 & 3.56 & 21.89 \\
        &&&& 26.16 && 130.46 && 123.97 && 626.27 & 2.43 & 14.84 \\
    BLG-501 & 13.31 & 12.24 & 24.1 & 33.0 & 89.9 & 123.6 & 317 & 253.3 & 1197.0 & 947.3 & 13.0 & 40.9 \\
    BLG-502 & 10.02 & 4.67 & 11.0 & 10.3 & 31.8 & 39.7 & 171 & 43.9 & 319.0 & 160.1 & 17.8 & 17.6 \\
    BLG-503 & 10.7 & 5.25 & 5.5 & 7.8 & 14.2 & 24.3 & 91 & 33.6 & 152.0 & 97.9 & 11.7 & 13.4 \\
    BLG-504 & 11.86 & 4.34 & 16.9 & 18.4 & 77.9 & 98.6 & 225 & 70.6 & 924.0 & 383.9 & 9.4 & 27.0 \\
    BLG-505 & 19.02 & 9.88 & 22.2 & 24.2 & 71.2 & 92.5 & 441 & 143.2 & 1354.0 & 550.1 & 9.5 & 37.1 \\
    BLG-506 & 12.85 & 5.36 & 16.5 & 20.7 & 57.0 & 88.3 & 247 & 81.7 & 732.0 & 357.5 & 9.5 & 20.1 \\
    BLG-507 & 11.22 & 5.27 & 12.3 & 15.4 & 33.3 & 59.3 & 216 & 67.1 & 374.0 & 259.3 & 17.0 & 14.2 \\
    BLG-508 & 9.53 & 5.0 & 6.8 & 10.4 & 22.9 & 33.4 & 119 & 45.4 & 218.0 & 145.1 & 9.4 & 14.4 \\
    BLG-509 & 9.82 & 4.72 & 4.6 & 6.1 & 13.3 & 20.6 & 79 & 23.8 & 131.0 & 79.0 & 11.0 & 7.8 \\
    BLG-510 & 7.67 & 3.47 & 3.6 & 3.0 & 6.6 & 10.2 & 38 & 7.8 & 51.0 & 27.9 & 6.4 & 4.1 \\
    BLG-511 & 13.45 & 5.11 & 13.5 & 16.8 & 56.1 & 64.0 & 204 & 64.0 & 755.0 & 244.2 & 4.5 & 23.1 \\
    BLG-512 & 17.48 & 8.34 & 14.0 & 18.0 & 53.0 & 63.9 & 280 & 108.4 & 926.0 & 384.9 & 9.9 & 25.2 \\
    BLG-513 & 13.71 & 7.87 & 8.5 & 14.6 & 26.8 & 54.0 & 213 & 78.1 & 368.0 & 287.5 & 8.7 & 14.1 \\
    BLG-514 & 10.27 & 5.44 & 6.2 & 10.2 & 18.9 & 41.8 & 108 & 45.9 & 194.0 & 179.9 & 9.8 & 8.0 \\
    BLG-515 & 8.1 & 4.32 & 5.7 & 8.1 & 19.6 & 24.8 & 80 & 38.3 & 159.0 & 115.1 & 1.7 & 10.7 \\
    BLG-516 & 7.61 & 3.76 & 4.6 & 7.0 & 7.6 & 15.0 & 48 & 17.8 & 58.0 & 41.0 & 0.0 & 1.5 \\
    BLG-517 & 5.96 & 3.18 & 2.6 & 4.5 & 0.3 & 11.3 & 12 & 9.6 & 2.0 & 23.1 & 2.1 & 0.2 \\
    BLG-518 & 12.34 & 4.72 & 7.3 & 11.1 & 25.2 & 36.1 & 160 & 38.9 & 311.0 & 126.7 & 4.4 & 12.6 \\
    BLG-519 & 12.93 & 6.3 & 9.3 & 10.5 & 29.3 & 36.1 & 223 & 51.4 & 379.0 & 177.3 & 11.3 & 21.1 \\
    BLG-520 & 11.63 & 5.94 & 5.6 & 9.1 & 20.3 & 32.1 & 124 & 54.5 & 236.0 & 187.7 & 13.8 & 9.2 \\
    BLG-521 & 9.18 & 4.95 & 4.5 & 9.2 & 12.2 & 26.4 & 69 & 40.9 & 112.0 & 127.0 & 0.0 & 7.8 \\
    BLG-522 & 8.62 & 4.24 & 7.4 & 7.8 & 19.4 & 21.7 & 108 & 40.5 & 167.0 & 107.8 & 10.4 & 12.9 \\
    BLG-523 & 10.4 & 4.84 & 6.0 & 7.0 & 16.5 & 21.1 & 99 & 26.6 & 172.0 & 76.3 & 10.0 & 9.8 \\
    BLG-524 & 9.66 & 4.49 & 6.0 & 6.5 & 7.9 & 20.2 & 75 & 17.0 & 76.0 & 50.7 & 9.4 & 4.9 \\
    BLG-525 & 7.84 & 4.04 & 3.2 & 4.3 & 6.5 & 15.4 & 36 & 16.4 & 51.0 & 55.1 & 7.9 & 5.6 \\
    BLG-526 & 5.25 & 3.34 & 2.3 & 5.1 & 4.4 & 13.5 & 20 & 34.0 & 23.0 & 88.8 & 3.2 & 1.7 \\
    BLG-527 & 6.36 & 2.85 & 5.5 & 2.89 & 5.3 & 8.30 & 48 & 10.63 & 34.0 & 33.64 & 8.44 & 1.42 \\
        &&&& 2.89 && 9.58 && 10.63 && 38.76 & 16.70 & 5.40 \\
        &&&& 3.06 && 9.55 && 11.29 && 38.76 & 8.03 & 1.24 \\
    BLG-528 & 7.39 & 2.99 & 5.5 & 4.1 & 5.7 & 11.2 & 50 & 12.8 & 42.0 & 33.1 & 0.0 & 0.6 \\
    BLG-529 & 7.79 & 3.2 & 5.6 & 4.7 & 3.6 & 11.9 & 39 & 7.0 & 28.0 & 15.3 & 1.9 & 1.8 \\
    BLG-530 & 7.37 & 3.24 & 8.5 & 2.4 & 4.9 & 9.9 & 54 & 3.5 & 36.0 & 14.2 & 10.6 & 2.7 \\
    BLG-531 & 6.85 & 3.05 & 5.1 & 3.0 & 4.8 & 9.5 & 34 & 6.4 & 33.0 & 20.9 & 6.9 & 5.0 \\
    BLG-532 & 6.02 & 2.82 & 2.7 & 2.6 & 2.0 & 6.9 & 23 & 8.2 & 12.0 & 24.7 & 4.8 & 1.4 \\
    BLG-533 & 1.02 & 0.21 & 7.1 & 0.0 & 0.0 & 0.0 & 1 & 0.0 & 0.0 & 0.0 & 0.0 & 0.0 \\
    BLG-534 & 9.06 & 4.51 & 17.2 & 19.8 & 72.3 & 101.1 & 176 & 93.2 & 655.0 & 481.0 & 3.6 & 19.3 \\
    BLG-535 & 7.57 & 3.58 & 15.2 & 14.4 & 43.7 & 67.1 & 189 & 46.0 & 331.0 & 217.5 & 9.4 & 9.0 \\
    BLG-536 & 5.21 & 2.47 & 3.8 & 1.9 & 0.8 & 4.8 & 15 & 0.7 & 4.0 & 1.7 & 2.1 & 0.6 \\
    BLG-539 & 4.52 & 2.37 & 4.0 & 2.8 & 0.4 & 6.1 & 13 & 8.3 & 2.0 & 18.4 & 3.9 & 0.4 \\
    BLG-543 & 4.76 & 1.62 & 5.9 & 2.9 & 7.1 & 10.8 & 36 & 5.0 & 34.0 & 18.9 & 9.3 & 2.8 \\
    BLG-544 & 5.66 & 2.87 & 6.1 & 5.2 & 6.0 & 15.0 & 40 & 11.7 & 34.0 & 33.2 & 0.0 & 1.9 \\
    BLG-545 & 6.47 & 3.05 & 6.1 & 6.4 & 14.1 & 17.6 & 57 & 14.8 & 91.0 & 40.1 & 7.7 & 8.0 \\
    BLG-546 & 4.25 & 2.47 & 3.4 & 3.8 & 0.5 & 7.7 & 12 & 7.5 & 2.0 & 14.6 & 0.0 & 0.2 \\
    BLG-547 & 3.67 & 2.07 & 1.3 & 1.99 & 0.3 & 5.46 & 3 & 2.57 & 1.0 & 7.80 & 0.88 & 0.19 \\
        &&&& 2.23 && 6.21 && 2.89 && 8.86 & 1.07 & 0.14 \\
        &&&& 1.74 && 4.47 && 2.25 && 6.38 & 0.00 & 0.18 \\
    BLG-566 & 3.03 & 1.28 & 0.9 & 2.4 & 0.0 & 4.0 & 2 & 5.0 & 0.0 & 7.8 & 0.3 & 0.0 \\
    BLG-573 & 7.16 & 3.83 & 3.7 & 8.1 & 4.9 & 19.5 & 33 & 14.5 & 35.0 & 35.3 & 4.9 & 3.5 \\
    BLG-580 & 8.47 & 3.19 & 8.0 & 9.2 & 29.2 & 31.9 & 116 & 22.6 & 247.0 & 76.7 & 7.2 & 9.6 \\
    BLG-588 & 5.39 & 3.19 & 2.7 & 6.5 & 3.9 & 12.4 & 23 & 13.6 & 21.0 & 27.9 & 3.6 & 2.4 \\
    BLG-597 & 4.28 & 2.57 & 4.0 & 2.2 & 3.7 & 7.4 & 27 & 5.3 & 16.0 & 17.5 & 2.9 & 1.9 \\
    BLG-599 & 7.8 & 3.69 & 3.9 & 4.5 & 3.8 & 11.3 & 41 & 14.6 & 30.0 & 34.5 & 5.0 & 2.4 \\
    BLG-600 & 5.29 & 2.5 & 2.9 & 3.5 & 2.3 & 6.0 & 18 & 9.0 & 12.0 & 14.5 & 1.4 & 0.8 \\
    BLG-603 & 10.02 & 5.24 & 10.6 & 11.1 & 29.8 & 36.6 & 167 & 50.2 & 299.0 & 168.8 & 5.8 & 19.5 \\
    BLG-604 & 9.8 & 4.42 & 7.0 & 6.4 & 12.4 & 20.3 & 100 & 19.7 & 122.0 & 65.6 & 7.0 & 9.1 \\
    BLG-605 & 6.13 & 2.6 & 3.9 & 1.8 & 4.2 & 8.1 & 29 & 3.9 & 26.0 & 15.8 & 12.8 & 3.8 \\
    BLG-606 & 4.48 & 2.32 & 3.8 & 1.8 & 0.2 & 3.8 & 11 & 2.8 & 1.0 & 6.2 & 1.6 & 0.1 \\
    BLG-609 & 6.77 & 3.05 & 8.4 & 7.9 & 21.6 & 36.6 & 81 & 17.9 & 146.0 & 92.5 & 5.2 & 6.4 \\
    BLG-610 & 5.78 & 2.86 & 6.2 & 5.9 & 9.3 & 19.6 & 38 & 12.3 & 54.0 & 42.3 & 0.0 & 1.9 \\
    BLG-611 & 6.93 & 5.14 & 16.2 & 18.79 & 68.0 & 75.07 & 158 & 154.64 & 471.0 & 624.8 & 8.34 & 23.25 \\
        &&&& 18.70 && 81.09 && 153.82 && 674.72 & 5.69 & 17.88 \\
        &&&& 16.88 && 78.52 && 139.01 && 653.92 & 7.27 & 20.63 \\
    BLG-612 & 7.34 & 4.14 & 10.1 & 14.3 & 21.0 & 56.5 & 87 & 30.2 & 154.0 & 117.9 & 4.1 & 5.2 \\
    BLG-613 & 7.99 & 3.96 & 12.4 & 15.1 & 24.8 & 68.0 & 119 & 40.4 & 198.0 & 181.3 & 11.4 & 11.1 \\
    BLG-614 & 6.32 & 3.33 & 5.5 & 10.4 & 13.3 & 31.7 & 43 & 23.5 & 84.0 & 72.2 & 3.0 & 5.4 \\
    BLG-615 & 8.09 & 3.7 & 6.8 & 13.5 & 19.0 & 40.8 & 74 & 31.0 & 154.0 & 94.1 & 4.8 & 5.6 \\
    BLG-616 & 5.08 & 2.75 & 3.4 & 3.7 & 0.2 & 11.4 & 12 & 5.5 & 1.0 & 14.5 & 1.5 & 0.1 \\
    BLG-617 & 6.32 & 3.08 & 5.2 & 6.20 & 10.6 & 16.75 & 43 & 18.79 & 67.0 & 50.63 & 5.45 & 6.64 \\
        &&&& 3.68 && 13.90 && 11.17 && 42.02 & 2.47 & 7.20 \\
        &&&& 5.19 && 14.57 && 15.75 && 44.05 & 3.43 & 3.64 \\
    BLG-619 & 4.47 & 2.57 & 3.3 & 6.8 & 0.0 & 19.6 & 3 & 1.9 & 0.0 & 5.2 & 0.7 & 0.0 \\
    BLG-621 & 5.15 & 2.24 & 6.8 & 5.1 & 8.3 & 11.0 & 44 & 11.5 & 43.0 & 22.9 & 8.4 & 4.1 \\
    BLG-622 & 4.76 & 2.43 & 3.6 & 5.3 & 6.7 & 15.7 & 25 & 13.0 & 32.0 & 35.4 & 3.6 & 3.1 \\
    BLG-624 & 6.02 & 2.54 & 4.3 & 3.0 & 6.3 & 8.7 & 31 & 6.7 & 38.0 & 21.0 & 2.9 & 4.0 \\
    BLG-625 & 6.47 & 2.6 & 5.5 & 5.1 & 7.9 & 15.4 & 46 & 11.6 & 51.0 & 30.7 & 10.6 & 5.3 \\
    BLG-626 & 8.32 & 4.3 & 8.3 & 9.2 & 15.9 & 32.9 & 92 & 31.1 & 132.0 & 112.8 & 7.3 & 4.2 \\
    BLG-629 & 4.56 & 2.09 & 3.4 & 2.71 & 0.4 & 7.65 & 11 & 4.71 & 2.0 & 12.75 & 3.9 & 0.37 \\
        &&&& 2.96 && 7.15 && 5.14 && 11.93 & 0.00 & 0.19 \\
        &&&& 2.96 && 7.15 && 5.14 && 11.93 & 1.85 & 0.24 \\
    BLG-630 & 5.11 & 2.21 & 4.4 & 5.4 & 7.4 & 11.2 & 30 & 11.9 & 38.0 & 25.5 & 8.4 & 5.4 \\
    BLG-631 & 7.09 & 2.99 & 7.1 & 6.9 & 8.7 & 20.3 & 61 & 16.2 & 62.0 & 47.6 & 8.9 & 3.5 \\
    BLG-632 & 9.59 & 5.43 & 9.0 & 10.2 & 22.8 & 33.1 & 128 & 45.8 & 219.0 & 155.7 & 5.6 & 10.2 \\
    BLG-633 & 6.3 & 2.78 & 13.1 & 12.2 & 23.3 & 55.5 & 106 & 27.4 & 147.0 & 119.9 & 4.8 & 6.1 \\
    BLG-636 & 4.26 & 2.26 & 5.1 & 5.5 & 0.5 & 11.9 & 14 & 8.1 & 2.0 & 16.5 & 0.9 & 0.1 \\
    BLG-637 & 5.84 & 2.56 & 5.2 & 7.4 & 8.6 & 19.5 & 40 & 14.9 & 50.0 & 41.4 & 1.6 & 2.0 \\
    BLG-638 & 7.96 & 3.66 & 6.9 & 10.1 & 11.1 & 32.2 & 64 & 22.4 & 88.0 & 73.9 & 5.1 & 3.7 \\
    BLG-639 & 5.08 & 1.89 & 10.4 & 8.2 & 19.5 & 34.8 & 70 & 13.4 & 99.0 & 54.2 & 3.2 & 3.7 \\
    BLG-641 & 5.33 & 3.05 & 4.9 & 9.6 & 4.5 & 25.8 & 31 & 22.7 & 24.0 & 60.0 & 2.4 & 1.2 \\
    BLG-642 & 3.22 & 1.37 & 8.7 & 4.1 & 8.7 & 15.8 & 36 & 5.0 & 28.0 & 19.6 & 0.0 & 1.3 \\
    BLG-643 & 2.11 & 1.53 & 5.7 & 4.7 & 8.5 & 16.1 & 17 & 8.4 & 18.0 & 29.8 & 0.0 & 0.7 \\
    BLG-644 & 1.02 & 0.34 & 0.0 & 0.0 & 0.0 & 0.0 & 0 & 0.0 & 0.0 & 0.0 & 0.0 & 0.0 \\
    BLG-645 & 3.33 & 0.66 & 13.6 & 5.4 & 32.1 & 13.9 & 63 & 3.4 & 107.0 & 9.4 & 0.0 & 5.6 \\
    BLG-646 & 5.09 & 0.62 & 8.3 & 8.3 & 19.8 & 19.9 & 57 & 4.6 & 101.0 & 9.5 & 0.0 & 0.0 \\
    BLG-647 & 0.68 & 0.21 & 0.0 & 0.0 & 0.0 & 2.4 & 0 & 0.0 & 0.0 & 0.1 & 0.0 & 0.0 \\
    BLG-648 & 2.86 & 1.74 & 18.3 & 8.0 & 42.3 & 45.5 & 67 & 12.1 & 121.0 & 67.7 & 6.9 & 4.5 \\
    BLG-649 & 0.93 & 0.18 & 0.0 & 0.0 & 0.0 & 0.0 & 0 & 0.0 & 0.0 & 0.0 & 0.0 & 0.0 \\
    BLG-650 & 0.91 & 0.18 & 0.0 & 0.0 & 0.0 & 0.0 & 0 & 0.0 & 0.0 & 0.0 & 0.0 & 0.0 \\
    BLG-651 & 1.56 & 0.77 & 0.0 & 8.0 & 0.0 & 20.8 & 0 & 1.1 & 0.0 & 2.9 & 0.0 & 0.0 \\
    BLG-652 & 6.55 & 3.62 & 14.2 & 16.2 & 22.3 & 84.5 & 100 & 33.9 & 146.0 & 177.9 & 7.3 & 5.0 \\
    BLG-653 & 5.37 & 4.56 & 20.1 & 20.1 & 39.9 & 109.4 & 127 & 64.1 & 214.0 & 362.3 & 4.7 & 5.0 \\
    BLG-654 & 4.92 & 2.35 & 16.7 & 15.4 & 48.8 & 84.9 & 110 & 27.8 & 240.0 & 159.4 & 5.6 & 9.0 \\
    BLG-655 & 2.13 & 0.25 & 9.5 & 0.0 & 0.0 & 2.1 & 3 & 0.0 & 0.0 & 0.1 & 0.0 & 0.0 \\
    BLG-657 & 4.43 & 2.38 & 2.6 & 1.9 & 0.7 & 7.1 & 6 & 3.4 & 3.0 & 11.0 & 0.9 & 0.5 \\
    BLG-659 & 1.05 & 0.73 & 0.0 & 8.5 & 0.0 & 21.1 & 0 & 0.9 & 0.0 & 1.6 & 0.0 & 0.0 \\
    BLG-660 & 4.63 & 2.07 & 23.7 & 12.2 & 35.4 & 50.6 & 94 & 15.7 & 164.0 & 72.4 & 10.6 & 10.6 \\
    BLG-661 & 4.22 & 2.78 & 15.2 & 7.6 & 25.6 & 44.6 & 80 & 14.7 & 108.0 & 88.9 & 0.0 & 5.5 \\
    BLG-662 & 6.16 & 1.79 & 12.2 & 12.1 & 20.0 & 56.0 & 87 & 12.8 & 123.0 & 66.2 & 10.4 & 7.7 \\
    BLG-663 & 2.15 & 1.35 & 3.7 & 6.9 & 2.3 & 23.3 & 4 & 6.6 & 5.0 & 22.4 & 0.0 & 0.0 \\
    BLG-664 & 1.85 & 0.64 & 5.7 & 1.6 & 4.3 & 8.8 & 6 & 1.1 & 8.0 & 3.1 & 2.1 & 0.7 \\
    BLG-665 & 2.72 & 1.34 & 1.2 & 5.8 & 0.4 & 16.9 & 1 & 1.2 & 1.0 & 3.3 & 0.1 & 0.0 \\
    BLG-666 & 2.04 & 1.9 & 9.8 & 7.6 & 0.0 & 34.6 & 3 & 2.1 & 0.0 & 12.3 & 0.3 & 0.0 \\
    BLG-667 & 2.41 & 1.72 & 11.3 & 11.4 & 15.8 & 43.8 & 27 & 17.8 & 38.0 & 61.7 & 2.4 & 1.3 \\
    BLG-668 & 1.93 & 0.48 & 8.8 & 4.3 & 5.2 & 10.8 & 15 & 1.4 & 10.0 & 3.7 & 0.0 & 0.0 \\
    BLG-670 & 1.65 & 1.5 & 10.7 & 4.5 & 10.9 & 14.8 & 19 & 3.8 & 18.0 & 21.5 & 0.0 & 1.2 \\
    BLG-672 & 2.45 & 2.88 & 5.6 & 5.36 & 8.6 & 17.14 & 17 & 12.03 & 21.0 & 34.56 & 3.05 & 1.43 \\
        &&&& 4.11 && 20.2 && 9.22 && 40.68 & 3.77 & 1.06 \\
        &&&& 5.36 && 18.6 && 12.03 && 37.44 & 3.05 & 2.49 \\
    BLG-675 & 5.64 & 5.5 & 26.5 & 22.2 & 53.2 & 118.5 & 160 & 91.9 & 300.0 & 491.8 & 9.3 & 9.8 \\
    BLG-676 & 4.58 & 4.27 & 3.9 & 21.7 & 0.0 & 136.9 & 1 & 5.9 & 0.0 & 38.3 & 0.1 & 0.0 \\
    BLG-677 & 0.96 & 0.84 & 0.0 & 5.5 & 0.0 & 15.9 & 0 & 0.3 & 0.0 & 1.1 & 0.0 & 0.0 \\
    BLG-680 & 3.57 & 2.0 & 5.0 & 3.3 & 0.8 & 11.9 & 8 & 1.8 & 3.0 & 9.5 & 1.2 & 0.3 \\
    BLG-683 & 4.28 & 3.51 & 26.2 & 16.0 & 1.6 & 74.2 & 21 & 7.6 & 7.0 & 37.7 & 1.2 & 0.3 \\
    BLG-705 & 1.1 & 0.65 & 0.0 & 0.0 & 0.0 & 0.0 & 0 & 0.0 & 0.0 & 0.0 & 0.0 & 0.0 \\
    BLG-706 & 0.94 & 0.54 & 3.4 & 0.0 & 0.0 & 0.0 & 1 & 0.0 & 0.0 & 0.0 & 1.0 & 0.0 \\
    BLG-707 & 0.81 & 0.5 & 0.0 & 0.0 & 0.0 & 1.0 & 0 & 0.0 & 0.0 & 0.5 & 0.0 & 0.0 \\
    BLG-708 & 0.84 & 0.42 & 0.0 & 1.2 & 0.0 & 1.2 & 0 & 0.5 & 0.0 & 0.5 & 0.0 & 0.0 \\
    BLG-709 & 0.61 & 0.0 & 0.0 & 0.0 & 0.0 & 0.0 & 0 & 0.0 & 0.0 & 0.0 & 0.0 & 0.0 \\
    BLG-710 & 0.59 & 0.0 & 2.6 & 0.0 & 0.0 & 0.0 & 1 & 0.0 & 0.0 & 0.0 & 0.0 & 0.0 \\
    BLG-711 & 0.48 & 0.0 & 0.0 & 0.0 & 0.0 & 0.0 & 0 & 0.0 & 0.0 & 0.0 & 0.0 & 0.0 \\
    BLG-714 & 5.84 & 2.84 & 11.0 & 11.9 & 7.4 & 57.2 & 54 & 14.8 & 43.0 & 74.8 & 5.3 & 1.6 \\
    BLG-715 & 3.42 & 1.61 & 10.2 & 6.1 & 14.9 & 30.1 & 36 & 8.0 & 51.0 & 31.1 & 8.0 & 3.5 \\
    BLG-717 & 2.06 & 0.77 & 3.9 & 2.7 & 1.0 & 8.6 & 6 & 1.6 & 2.0 & 3.5 & 1.7 & 0.1 \\  
     \cline{1-13}
\enddata
\tablecomments{\change{Microlensing estimates for the following, all scaled to the full OGLE-IV field area of 1.4 $\text{deg}^2$: (1) ``Mr\'{o}z19": observational results from \citet{Mroz:2019}, (2) ``M": our ``Mr\'{o}z19 Cut" simulation results, (3) ``EWS": Early Warning System events triggered between the 2011 and 2018 OGLE-IV seasons, and (4) ``E": our ``EWS Cut" simulation results. $N_{s}$ is the completeness corrected number of stars brighter than the magnitude limit of the survey ($\color{black}I_{V}=21$), $\Gamma$ is the completeness corrected event rate (excluding PBH events), $N_{\text{ev}}$ is the appropriately scaled number of events excluding PBHs (using Equation \ref{eq:ev Mroz}), and $N_{\text{ev, PBH}}$ is the number of scaled simulated PBH microlensing events (using Equation \ref{eq:PBH Mroz}). Columns with three values were run over three seeds each, while only one seed is run for each remaining OGLE-IV estimates. Poisson uncertainty is assumed for all simulated values.}}
\end{deluxetable*}

\begin{figure*}[t]
    \begin{center}
        \includegraphics[width=.86\textwidth]{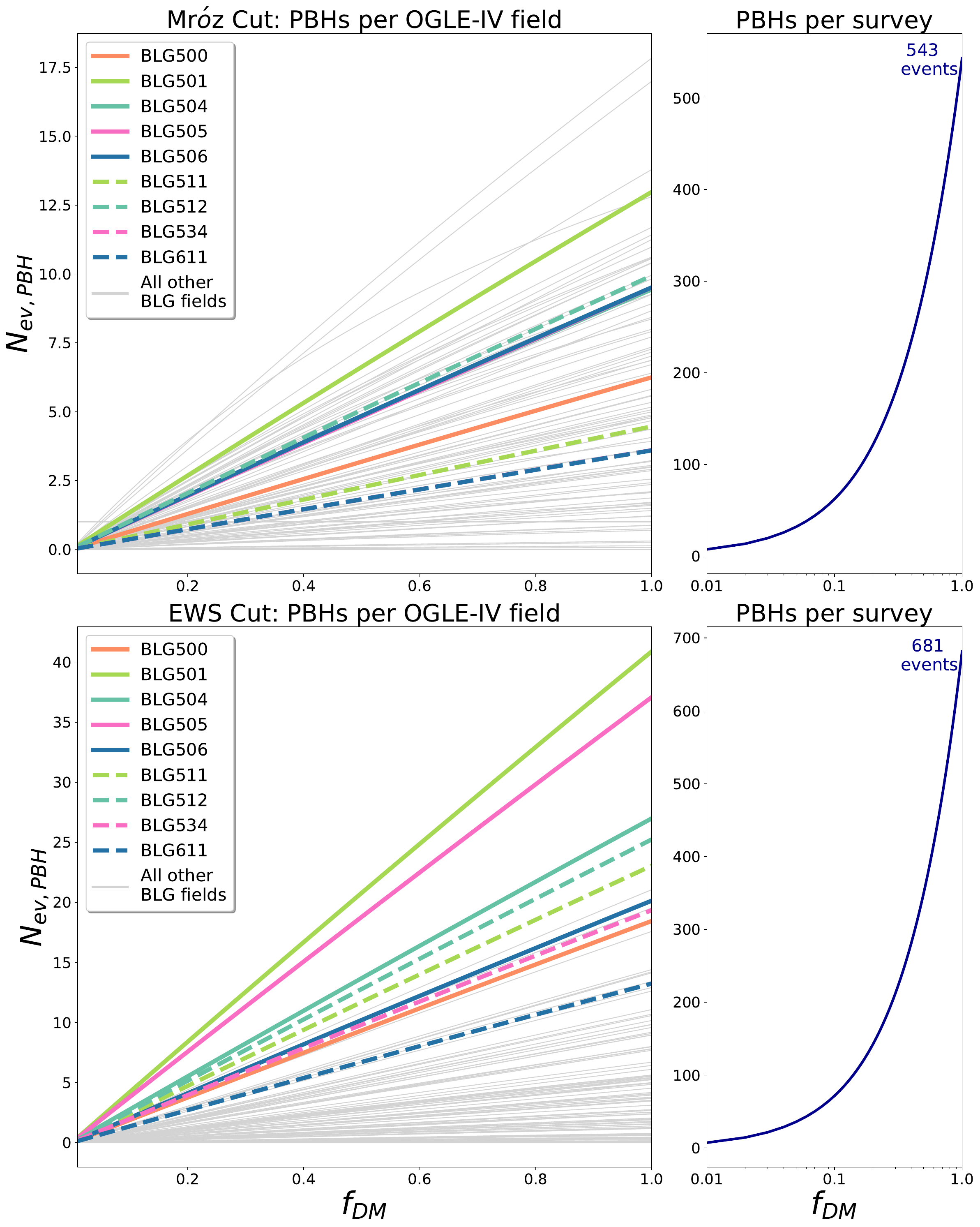}
        \caption{\change{OGLE-IV PBH microlensing event ($N_{\text{ev, PBH}}$) estimation as a function of dark matter fraction ($f_\text{DM}$). \emph{Upper Left:} PBH event estimation per OGLE-IV field, for our Mock Mr\'{o}z19 simulations. High cadence and low cadence OGLE-IV BLG fields are shown with bold colored lines, and thin gray lines, respectively. \emph{Upper Right:} Cumulative PBH event estimate over all fields, with the $f_\text{DM}$=1 estimate reported. \emph{Lower Left:} same as upper left but for our EWS cut. \emph{Lower Right:} same as upper right but for our EWS cut.} \update{$N_{\text{ev, PBH}}$ has a non-linear dependence on $f_{\text{DM}}$ (see Equation \ref{eq:PBH Mroz}), meaning lines on the left panel are not always linear.}}
        \label{fig:num_pbh_ogle}
    \end{center}
\end{figure*}

\change{Table \ref{tab:OGLE Comparison All} assumes that 100\% of dark matter consists of PBHs. Figure \ref{fig:num_pbh_ogle} further illustrates the breakdown of the number of predicted PBH events per field, as a function of $f_{\text{DM}}$. The high-cadence OGLE-IV fields (from \citet{Mroz:2019}, labeled in bold on the left panel of Figure \ref{fig:PopSyCLE fields}) are plotted in color on Figure \ref{fig:num_pbh_ogle}, while each low-cadence field is plotted as a gray line. The right panels of Figure \ref{fig:num_pbh_ogle} show the cumulative number of PBH events over each field, giving the total number of estimated events per survey. In all panels, the number of PBH events is determined using Equation \ref{eq:PBH Mroz}.} \update{Because Equation \ref{eq:PBH Mroz} has a factor of $f_{\text{DM}}$ in each instance of $N^{sim}_{\text{ev, PBH}}$, there is a non-linear dependence on $f_{\text{DM}}$ in $N_{\text{ev, PBH}}$, making some lines on Figure \ref{fig:num_pbh_ogle} curved.}

We combine results across all simulated OGLE\hyp{}IV fields with the EWS cut and plot the resulting $\pi_{E}$, $t_E$ distribution for all events (see left panel of Figure \ref{fig:pi_E_t_E_dist_all}). We then compare the $t_E$ distributions between PBHs and all populations except PBHs, and then compare the fractional contributions between compact object populations. The top panel of Figure \ref{fig:frac} shows our simulated EWS $t_E$ results. The upper row shows the $t_E$ distributions for all populations combined, all populations except PBHs, and PBHs alone, while the bottom row shows the relative contribution of each object type to the total number of microlensing events, as a function of $t_E$. The columns of each panel correspond to $f_{\text{DM}}=0.01$, 0.1, and 1, respectively. \change{Both Figure \ref{fig:pi_E_t_E_dist_all} and Figure \ref{fig:frac} include values from one seed run per field, and are generated before applying any of the scaling factors given in Equations \ref{eq:ev Mroz} or \ref{eq:PBH Mroz} (i.e., these are the total subset of events output from \texttt{PopSyCLE} with only the cuts listed in Table \ref{tab:Survey Parameters}).}

\subsection{Mock Roman Simulations}
\label{sec:roman sims}

The Nancy Grace Roman Space Telescope \citep[abbreviated ``Roman" and previously named WFIRST;][]{Spergel:2015}, is a space\hyp{}based telescope set to launch in 2027. During Roman's $\sim$5 year operation, Roman will conduct a microlensing survey towards the galactic bulge consisting of a seven\hyp{}field mosaic, over six 72 day seasons \citep{Penny:2019, Johnson:2020}. During each microlensing survey season, images are estimated to be taken every $\sim$15 minutes from one of the seven fields in the ``Cycle 7" configuration \citep[see Figure 7 of][]{Penny:2019, Johnson:2020}, the survey footprint of which spans roughly the region between $l \approx (-0.5, 1.5)$ and $b \approx (-0.5, -2)$).

\begin{figure*}[t]
    \begin{center}
        \includegraphics[width=.99\textwidth]{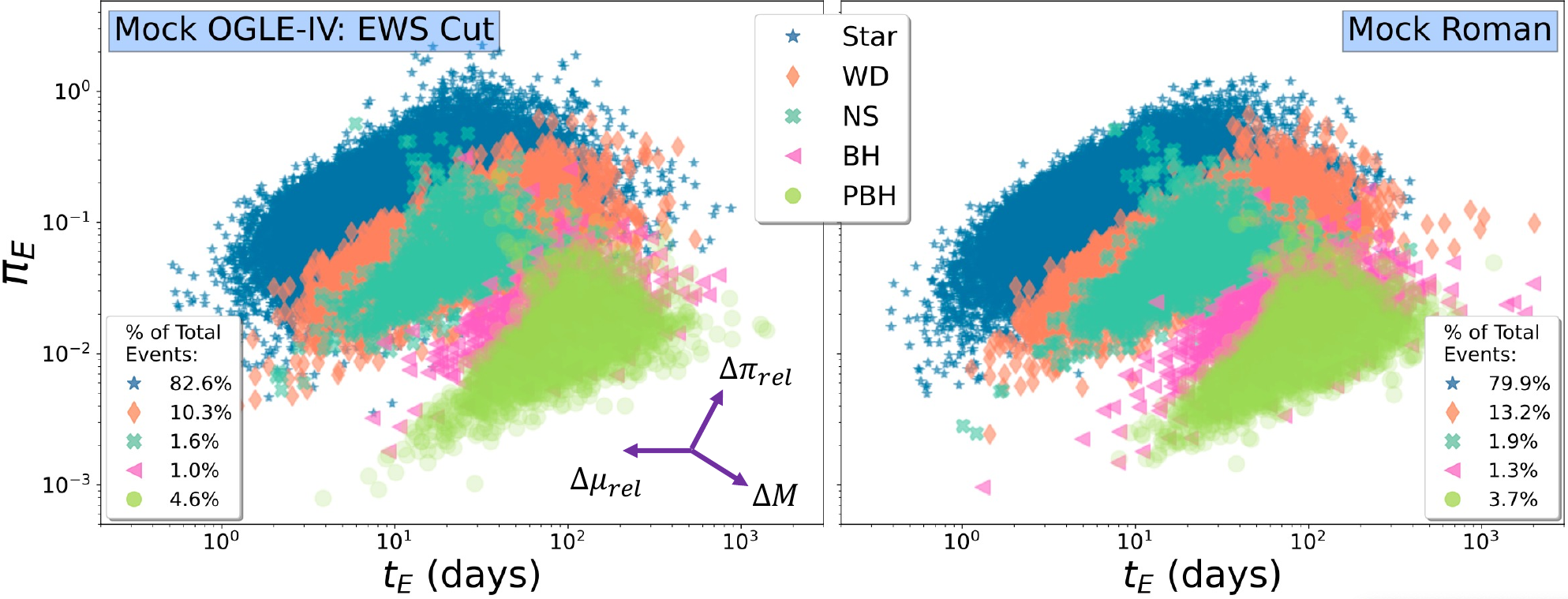}
        \caption{Simulated microlensing parallax ($\pi_{E}$) and Einstein crossing time ($t_E$) distributions. \emph{Left}: Mock OGLE\hyp{}IV simulation distribution after applying the Mock EWS observational cuts from Table \ref{tab:Survey Parameters}. \change{One seed run from each field is represented on this panel.} \emph{Right}: Same as left panel, this time for Mock Roman. The arrows on the bottom of the left panel demonstrate the direction that points on this plot would shift, as a function of increasing the quantities specified. Only a small portion of the Roman survey footprint is simulated, and therefore only our simulated regions are represented within the right panel. \change{This panel reflects events from all three seed runs per field. Neither panel considers any scaling or correction factors, and are considered the number of \texttt{PopSyCLE} detections given only the cuts from Table \ref{tab:Survey Parameters}. This figure should be used for distribution demonstration purposes only.}}
        \label{fig:pi_E_t_E_dist_all}
    \end{center}
\end{figure*}

Our Mock Roman survey uses the corresponding values from Table \ref{tab:Survey Parameters} to simulate the Mock Roman labeled fields in Figure \ref{fig:PopSyCLE fields}, which cover a subset of the projected Roman microlensing survey footprint. The values in this table match the ``Mock WFIRST" values from Lam\citeyear{Lam:2020}, which come from a combination of \citet{Penny:2019} and OGLE EWS values, and are described by Lam\citeyear{Lam:2020} as ``reasonable criteria expected for the mission". Each field is simulated three times, each with a different random seed, with all calculated quantities reported in this work using the average of all simulated results from that field. We assume constant stellar density over each simulated field whenever interpolating results.

\begin{figure*}[p]
    \centering
        \includegraphics[width=.885\textwidth]{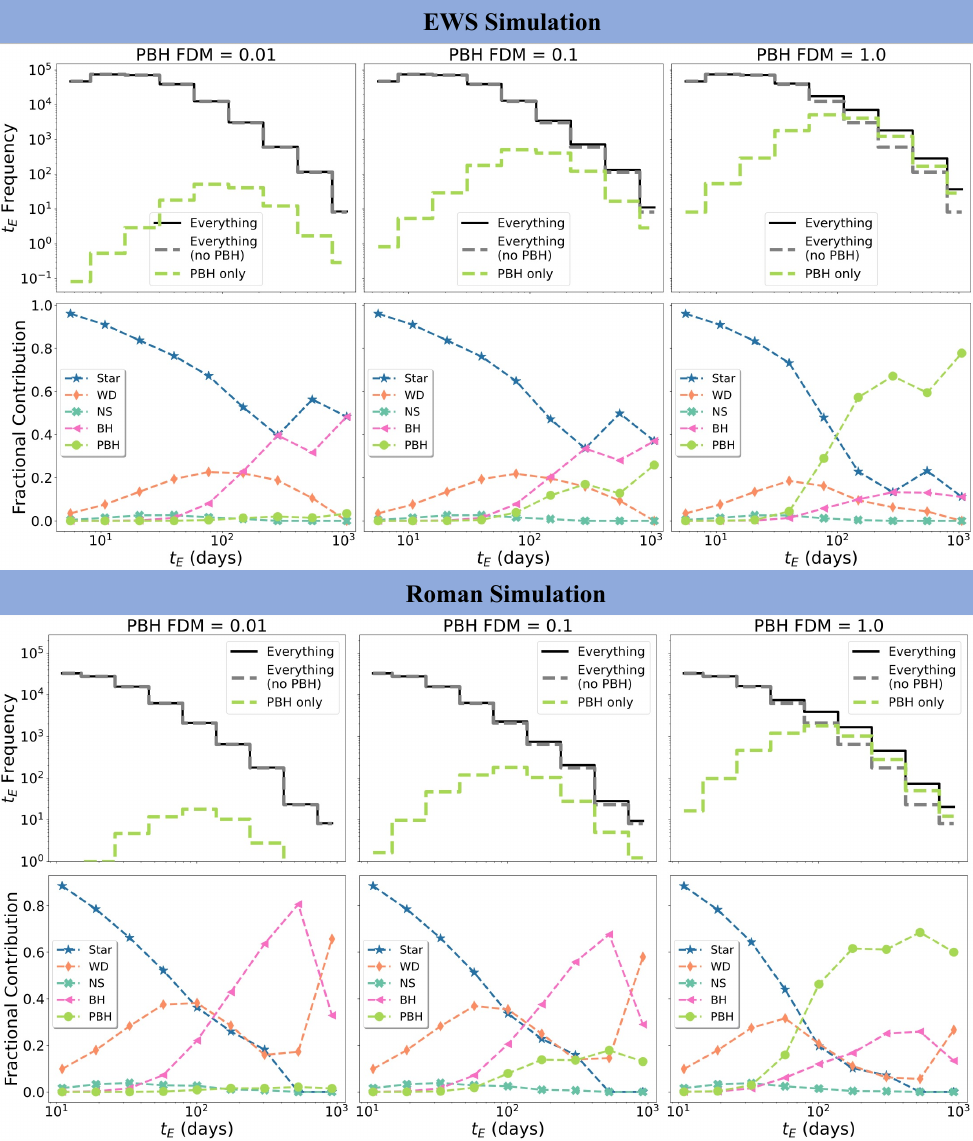}
        \caption{$t_E$ distributions and fractional contributions across various compact object populations. \emph{Upper}: Mock OGLE\hyp{}IV simulations (EWS cut) results\change{, from one seed per field}. \emph{Lower}: Mock Roman survey results\change{, including all three seed runs for each of the three simulated Roman fields.} The left, middle, and right columns of each panel correspond to $f_{\text{DM}}=1$, $f_{\text{DM}}=0.1$, and $f_{\text{DM}}=0.01$, respectively. The upper row of each panel has three $t_E$ distributions: all compact object populations, all populations except PBHs, and only the PBH population. Median PBH $t_E$ values are shown by the dotted vertical lines. The lower row of each panel shows the fractional $t_E$ contribution from each population, as a function of Einstein crossing time. \change{This figure should be used for distribution demonstration purposes only.}}
        \label{fig:frac}
\end{figure*}

The Roman microlensing survey will be divided up into seasons throughout Roman operation (exact dates are not yet confirmed), with an expected operational timeline of $\sim$5 years. Due to the microlensing survey seasons not being fully defined, we choose to spread the total microlensing survey over the estimated operational time, then scale observing frequency accordingly. Given the six planned seasons, each 72 days in length, with $\sim$96 observations each day, we estimate 41,472 observations total. Then, spreading the number of observations equally over the 5 year estimated timeline works out to $\sim$23 observations per day. Because \texttt{PopSyCLE} cadence does not account for various time gaps, we evenly spread the observations over the telescope duration, ignoring daytime gaps and survey gaps. Because we do not expect any PBH events with $t_E < 10$ days, and we want to improve our computational feasibility, we choose to slow the cadence to 1 observation every 3 days. This enables us to stay within our computational boundaries, while still accurately representing the $t_E$ distribution peak for all other compact objects.

Slowing the cadence this way can lead to underestimating the number of potentially detectable events, however, our choice of cadence should not affect the BH and PBH populations. Additionally, the actual Roman survey layout (i.e. seasons throughout telescope operation) will lead to event overestimation, as some events will fall entirely within observing gaps between seasons, or will be missed by only detecting one side of the light curve, etc. The actual Roman survey may be more sensitive to short duration events, but fail to detect a fraction of events with timescales that fall between seasons (which can be $>$ 200 days if seasons are evenly spaced out over operation), which are detected in our \texttt{PopSyCLE} simulations. We estimate these gross cadence and duration approximations to be sufficient for an order of magnitude estimate of the number of PBH microlensing events expected with Roman.


The right panel of Figure \ref{fig:pi_E_t_E_dist_all} illustrates the $\pi_{E}$, $t_E$ distribution for all events from all object populations, and the lower panel of Figure \ref{fig:frac} shows the $t_E$ distributions and fractional $t_E$ contribution across populations for our Mock Roman simulations. \change{Note that this plot reflects events from all three seed runs, from only a portion of the full Roman survey footprint, and has no scaling applied beyond the Table \ref{tab:Survey Parameters} cuts, as they are strictly for distribution demonstration.}

\begin{figure*}[tb]
    \begin{center}
        \includegraphics[width=.99\textwidth]{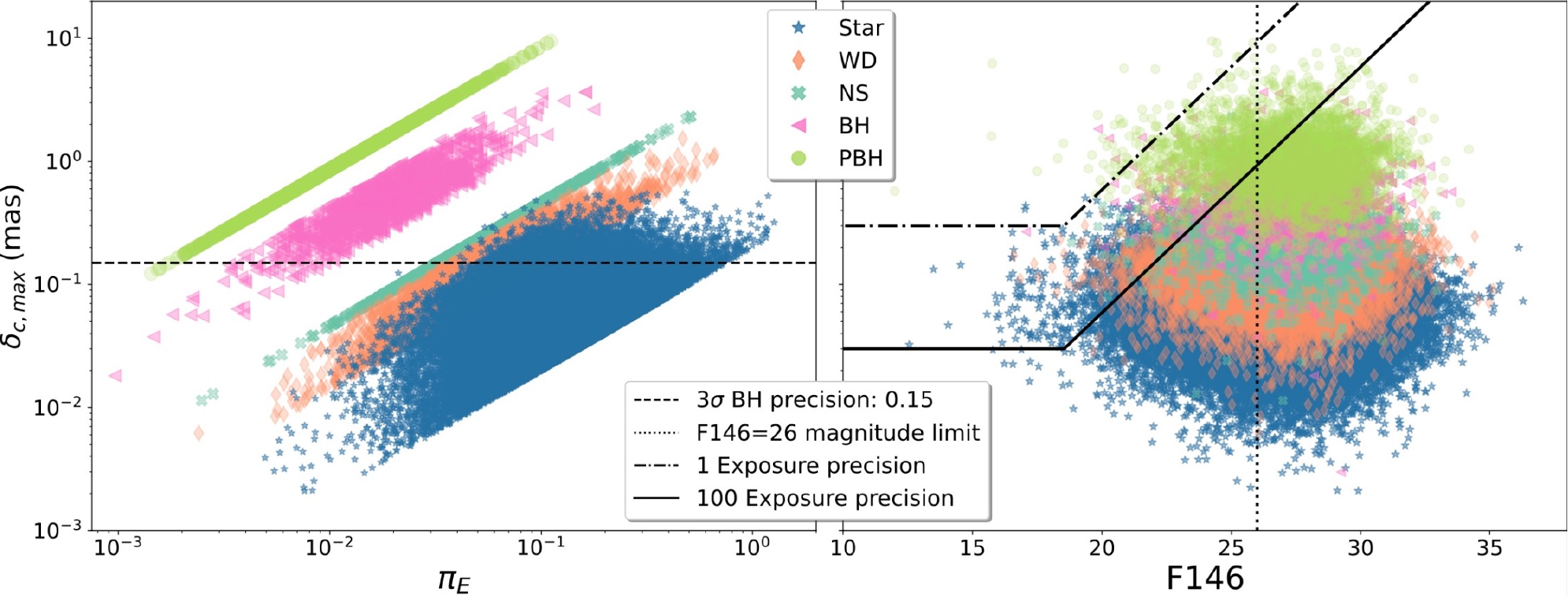}
        \caption{Simulated Roman survey distributions\update{, after cuts from Table \ref{tab:Survey Parameters}.} \emph{Left:} Maximum astrometric shift ($\delta_{c, max}$) vs. microlensing parallax ($\pi_E$). \emph{Right:} Maximum astrometric shift ($\delta_{c, max}$) vs. \update{F146 AB} source magnitude. The horizontal dashed line represents the 3$\sigma$ astrometric precision required for Roman ($\sim$0.15 mas) to detect isolated black holes \citep{sanderson:2019}, \change{the vertical dotted line represents the approximate limiting magnitude for Roman (\update{F146 $\approx$ }26), and the two diagonal lines represent the 1 exposure (dash-dotted line) and 100 exposure (solid line) estimated precision curves. Roman should be able to detect events that fall above the solid black line, and that fall below \update{(i.e., to the left of)} the limiting magnitude of the survey. Note that this plot includes events from simulations on only a small portion of the full planned survey footprint, all three seed runs per simulated field are included in this figure, and events represented here include only the cuts from Table \ref{tab:Survey Parameters}.}}
        \label{fig:pi_e_delta_c_dist_all}
    \end{center}
\end{figure*}

Due to its resolving power and diffraction limited seeing, Roman will be capable of obtaining simultaneous photometric and astrometric measurements from observed microlensing events, with an estimated astrometric precision (for isolated BHs) of $\sim$ 0.05 $mas$ \citep{sanderson:2019}. \change{Blending between the source and lens needs to be considered for luminous lenses, but this work primarily examines dark lenses, thus we do not account for source-lens blending. We also choose to ignore blending from neighboring stars as we expect the scatter in including this blending to be sub-dominate to other simplifying assumptions. Therefore, we define the maximum astrometric shift as $\delta_{c, max}=\delta_{c}$ (where $\delta_{c}$ is given in Equation \ref{eq:astro shift}), calculated at the point in which the maximum astrometric shift occurs, $u = \pm{\sqrt{2}}$}. This yields the maximum distance that the image will deviate from the expected path of the star with no lensing present. \change{The left panel of Figure \ref{fig:pi_e_delta_c_dist_all} illustrates $\delta_{c, max}$ as a function of $\pi_E$, and includes a dashed line for the 3$\sigma$ astrometric precision required for Roman ($\sim$0.15 $mas$) to detect isolated black holes.}

Additionally, due to the assumed delta\hyp{}function mass distribution for PBHs and NSs, they appear highly correlated in this space. However, extending the mass distributions will extend their footprint in this space, likely causing overlap with the BH population. \change{Resulting $\delta_{c, max}$ vs. $\pi_E$ distributions and $\delta_{c, max}$ vs. \update{F146 AB} source magnitude distributions for our Mock Roman simulation are shown on the left and right panels of Figure \ref{fig:pi_e_delta_c_dist_all}, respectively.}

\change{We estimate the number of PBH microlensing events that may be observed by Roman in three ways:}

\begin{enumerate}
    \item \change{\update{Less} conservative cut: estimate the number of PBH events in each of our simulated 0.16 $\text{deg}^2$ fields by first averaging over the three seed runs per field, then scale the results up from $3\times0.16$ $\text{deg}^2$ to the full Roman survey area of 1.97 $\text{deg}^2$}
    \item \change{\update{More} conservative cut: multiply the result from method 1 by the survey duty cycle:}
    \begin{equation}
        \text{duty cycle} = \frac{6 \text{ seasons} \times 72 \text{ days}}{5 \text{ years} \times 365.24 \frac{\text{ days}}{\text{ year}}} = 0.24
        \label{eq:duty cycle}
    \end{equation}
    \item \change{Realistic cut: count up the number of events that fall both above the solid 100 exposure precision line on Figure \ref{fig:pi_e_delta_c_dist_all} and to the left of the dotted $\color{black}F146=26$ limiting magnitude limit line, divide results by three (as each of the three seeds is represented on that plot), then scale up from $3\times0.16$ $\text{deg}^2$ to the full Roman survey area of 1.97 $\text{deg}^2$}
\end{enumerate}

\update{To determine our realistic event estimate cut given survey magnitude and astrometric shift constraints, we first apply the observational cuts from Table \ref{tab:Survey Parameters}. For the initial magnitude cut we remove events where baseline magnitude (i.e., combined flux from source, lens, and neighbors) is $H_{AB}>$ \maybe{26.7} \citep{sanderson:2019}\footnote{\update{$H_{AB}=$ \maybe{26.7} is the Roman High-Latitude Survey limit for faint point-sources \citep{sanderson:2019}. We assume this value to be an appropriate initial cut.}}. We then use Equation 1 from \citet{wilson:2023} to convert $H_{AB}$, $J_{AB}$, and $K_{AB}$ into the Roman F146 AB magnitudes. Then, we use a modified version of Equation 4 from \citet{Fardeen:2023} to estimate the astrometric detection threshold of Roman as a function of magnitude, given a single exposure:}

\begin{equation}
    \color{black}\sigma_{\text{ast}} = \text{MAX}(0.3, 10^{0.2 \times F146-4.23})
    \label{eq:fardeen}
\end{equation}
\update{\citet{Fardeen:2023} uses a minimum astrometric precision of 0.1 mas, however, that is the optimal Roman detection case. Our modified equation uses 0.3 mas as the minimum astrometric precision, which is the minimum value for Roman to ``properly sample astrometry" at the saturation limit of $H_{V}$ $\approx$ 18 \citep{Lam:2023}:}

\update{For the 100 exposure limit curve (the solid black line on the right panel of Figure \ref{fig:pi_e_delta_c_dist_all}), we assume $\sigma_{\text{ast}}$ $\propto$ $1/\sqrt{100}$. We note that this SNR calculation neglects the negative effect that blending will have on astrometric precision, as it leads to higher correlated noise and generally smaller astrometric signals, and is therefore overly optimistic when estimating the predicted number of astrometric events\footnote{\update{For Roman, when the magnitude $F146 < 26$, the density of sources is expected to be $\approx$ 10 stars per arcsecond \citep{Penny:2019}, which will make precision astrometry challenging.}}}.

\change{The right panel of \ref{fig:pi_e_delta_c_dist_all} shows maximum astrometric shift, $\delta_{c, max}$, vs. \update{F146 AB} source magnitude, for each of our individual microlensing events. Also marked on this figure is Roman's estimated limiting magnitude\update{, F146 $\approx$ }26). Roman should be able to detect events that lie above the solid black line, and that fall below the limiting magnitude of the survey (which we estimate here as \update{falling to the left of} the dashed line at \update{F146}=26). Figure \ref{fig:pi_e_delta_c_dist_all} includes the full distribution of events over all three seeds, over all three simulated Roman fields, and assume not scaling ratios or cuts other than those from Table \ref{tab:Survey Parameters}.}

\section{Results}
\label{sec:Results}

Adding a PBH model to \texttt{PopSyCLE} serves as a capable means for simulating PBH dark matter, enabling us to estimate the number and statistical distribution of detectable PBH microlensing events for various microlensing surveys, and to constrain the abundance of PBH dark matter in the Milky Way. Throughout the remainder of this section we discuss the results from our Mock OGLE\hyp{}IV and Mock Roman simulations, comparing estimated photometric and astrometric measurement distributions between PBHs and other compact object populations, using similar analysis techniques as used by Lam\citeyear{Lam:2020}.

Lam\citeyear{Lam:2020} find that analyzing microlensing events in $\pi_{E}$, $t_E$ space proves to be a compelling way of disentangling black holes from other types of lenses, given only photometric information. While $\pi_{E}$, $t_E$ space can enable ensemble statistical constraints, it does have real\hyp{}time alert limitations. Lam\citeyear{Lam:2020} notes that this analysis cannot be criteria for astrometric follow\hyp{}up selection, as $\pi_{E}$ is difficult to measure precisely at or before the time of the photometric peak of the light curve occurs. Fortunately, the next wave of microlensing surveys (e.g., Roman) are expected to obtain simultaneous photometric and astrometric observations, and Lam\citeyear{Lam:2020} further finds that combining photometric analysis with astrometric measurements has the ability to further increase separability between populations, making \texttt{PopSyCLE} a powerful tool for potentially differentiating PBHs from other compact objects in these parameter spaces.

\subsection{Mock OGLE\hyp{}IV}
\label{sec:ogle}
Given the resulting PBH event distributions from Figure \ref{fig:pi_E_t_E_dist_all}, we find that 30 $M_{\odot}$ PBH events tend towards higher $t_E$ and lower $\pi_E$, but that the population significantly overlaps the stellar evolved BH distribution. This overlap is likely due to our choice of PBH mass, as we would only expect PBHs to shift in this space by a factor of $1/\sqrt{\bar{m}_{\text{BH}}}$ in $\pi_E$, and by a factor of $\sqrt{\bar{m}_{\text{BH}}}$ in $t_E$. Additionally, the median $\mu_{rel}$ is slightly higher for PBHs than BHs, with $\mu_{rel, \text{PBH}}$ $\sim$ \change{12.5} and $\mu_{rel, \text{BH}}$ $\sim$ \change{7.1} (averaged between the Mr\'{o}z19 and EWS cuts), causing an additional shift to lower $t_E$, undoing some of the offset shift to higher $t_E$ due to the higher PBH masses.


Comparing the $t_E$ distributions with and without PBHs in Figure \ref{fig:frac} reveals that PBH microlensing events peak at longer duration $t_E$ than other compact objects, with a peak of $\sim$ \change{91.7} days for both our Mr\'{o}z19 and EWS cuts. Additionally, analyzing the fractional contribution of PBH to BH events in our EWS cut simulations \change{(using the full distribution of events after only the cuts from Table \ref{tab:Survey Parameters})}, we find $\sim$ \change{4.3}$f_{\text{DM}}$ times more PBH events than BH events. Note that our results use a higher cadence than the Lam\citeyear{Lam:2020} ``v3" results, catching more events with long duration. However, when running our simulations without PBHs using the exact same simulation parameters as Lam\citeyear{Lam:2020}, we find identical results to those reported in Lam\citeyear{Lam:2020} Appendix A.1 and by \citet{Rose:2022}. 

\change{We find a small shift in Einstein crossing time peak values between our simulations and \citet{Mroz:2019} \citep[consistant with ][]{Rose:2022}, likely due to field-of-view extrapolation and the absence of binary lenses and binary sources, which should not have a significant impact on our predicted long-duration PBH events. Similar to \citet{Rose:2022} and Lam\citeyear{Lam:2020}, our star counts and event rates can be off from \citet{Mroz:2019} and EWS estimates by a factor of $\approx$ 2 (see Table \ref{tab:OGLE Comparison All}). The star count differences are not due to stellar binarity or confusion \citep[Abrams et al. in prep., ][]{Rose:2022}, and the event rate discrepancies may be due to the galactic model chosen for our simulations and extinction being variable closer to the galactic bulge (Lam\citeyear{Lam:2020}). \change{It is worth noting that our EWS predicted PBH event rates will be artificially low, as we have not accounted for spurious and anomalous events, beyond our general scaling factor between our simulated results without PBHs vs. the observational EWS results.}}

\change{Given Table \ref{tab:OGLE Comparison All} results, and Figure \ref{fig:num_pbh_ogle}, we find that 681 PBH events may be within the 8 year OGLE-IV dataset, in which each PBH may conceivably be detected, but reliable lensing parameter estimates may not be attainable for (i.e., EWS cuts). We additionally estimate 543 PBH events may be detectable and also have sufficient SNR such that lensing estimates can be reasonably estimated for (i.e., Mr\'{o}z19 cuts).}

\subsection{Mock Roman}
\label{sec:roman}

Figure \ref{fig:pi_E_t_E_dist_all} illustrates the distribution results that we expect to be detectable in the upcoming Roman microlensing survey. We find a significant overlap between the PBH and BH populations which we expect to be for the same reason as described in \S \ref{sec:ogle}, which is mainly due to our choice of $\bar{m}_{\text{PBH}}$ and the resulting $\mu_{rel}$ distributions, with median values of $\mu_{rel, \text{PBH}}$ $\sim$ \maybe{11.3} and $\mu_{rel, \text{BH}}$ $\sim$ \maybe{7.0}. We find $\sim$ \maybe{2.8} $f_{\text{DM}}$ times more PBH microlensing events than we do BH microlensing events, and find a peak $t_E$ of $\sim$ \maybe{91.3} days. The estimated number of PBH events to BH events is lower than that of our simulated OGLE survey, but that is expected, as PBH events tend towards longer timescale events, and OGLE is a longer survey.

Given our simplified PBH model and assumed Roman microlensing survey structure, we estimate the number of potentially detectable PBH microlensing events expected with Roman using the three calculation methods outlined in Section \ref{sec:roman sims}. For scaling up from our simulated survey area of 0.48 $\text{deg}^2$ to the full Roman survey area of 1.97 $\text{deg}^2$, we account for the overlap between two of our simulated fields (where the overlap fraction = 0.19\%). We assume constant stellar density over the Roman microlensing survey footprint, $\bar{m}_{\text{PBH}}=30$ $M_{\odot}$, and that there are no survey gaps. \change{For our \update{less} conservative event estimate (scaling method 1, that scales our predicted event rate up to the full survey area of 1.97 $\text{deg}^2$), we estimate \maybe{4,981} events to be detectable with Roman. Our \update{more} conservative estimate (scaling \update{method 2}, which multiplies the result of \update{scaling method} 1 by the Roman duty cycle given in equation \ref{eq:duty cycle}, and then scales that result up to the full survey area of 1.97 $\text{deg}^2$) results in \maybe{1,195} PBH events that might be detectable by the future Roman survey. \update{Lastly}, our realistic cut estimate (scaling method \update{3}, which includes only the events that would fall above the solid 100 exposure precision line, and to the left of the limiting magnitude dashed line on Figure \ref{fig:pi_e_delta_c_dist_all}, and scaled to the full survey size) results in \maybe{1,212} PBH events detectable by Roman.}

Ignoring potentially important survey selection effects, the actual number of expected PBH lensing events will scale roughly proportional to $f_{\text{DM}}$ and approximately constant as a function of mass. Because we slow the cadence from $\sim$ 23 observations a day to 1 observation every 3 days (see calculations in \S \ref{sec:roman sims}), we underestimate the signal\hyp{}to\hyp{}noise of microlensing events by a factor of $\sim$ 8, and ultimately underestimate the number of microlensing events that may be detectable. However, we then overestimate the predicted number due to our survey gap assumptions\update{, and assumption of a source-dominated noise regime}. \change{For these reasons, we expect our event estimates to \update{conservatively} hold to $\sim$\update{1,000}$f_{\text{DM}}$.}

\begin{figure}[b]
    \begin{center}
        \includegraphics[width=.49\textwidth]{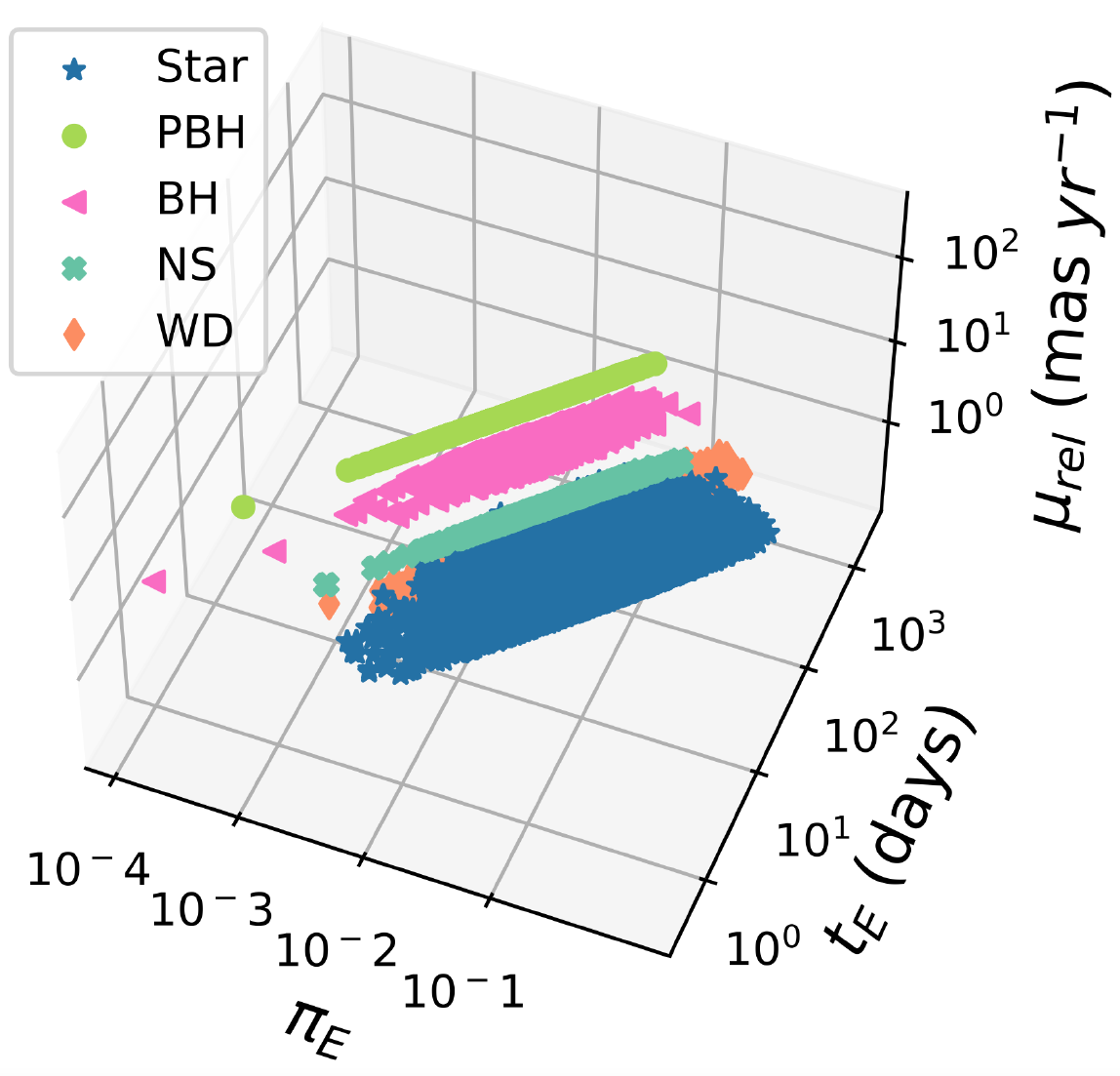}
        \caption{$\pi_E$, $t_E$, $\mu_{rel}$ distribution of compact objects for our simulated Roman survey (using all three seeds from each of the three simulated fields\change{, including only the cuts from Table \ref{tab:Survey Parameters})}. This figure demonstrates that while there is significant overlap of the BH and PBH populations in $\pi_E$, $t_E$ space (Figure \ref{fig:pi_E_t_E_dist_all}), there is potential to disentangle the populations with the addition of a $\mu_{rel}$ component.}
        \label{fig:pi_e_t_e_mu_rel}
    \end{center}
\end{figure}

Analyzing event distributions in $\delta_{c, max}$, $\pi_E$ space (left panel of Figure \ref{fig:pi_e_delta_c_dist_all}) finds separability between the PBH and BH populations, more so than in $\pi_E$, $t_E$ space. Note that because we use a delta\hyp{}function PBH mass distribution, the astrometric shift appears as a perfect linear function in this space, however, exploring broader mass spectra will cause scatter about $\delta_{c, max}$, likely causing overlap with the BH population. We then explore the compact object event distributions in three-dimensional $\pi_E$, $t_E$, $\mu_{rel}$ space, and find complete separability between the PBH and BH populations (Figure \ref{fig:pi_e_t_e_mu_rel}). Again, this complete separability is likely due to our choice of a constant mass, however, more analysis is needed to confirm how much overlap would appear in this space given an extended PBH mass distribution.

In general, decreasing $\bar{m}_{\text{PBH}}$ will result in PBH microlensing events overlapping in $\pi_E$, $t_E$ space with compact object microlensing events with similar masses, although different spatial and velocity distributions will not make this exact. Additionally, our simulations assume no blending, however, due to the sensitivity of Roman, binary sources and flux from neighboring background stars will dilute the astrometric signal, causing Figure \ref{fig:pi_e_delta_c_dist_all} and Figure \ref{fig:pi_e_t_e_mu_rel} to change, introducing scatter in the $\delta_{c, max}$ and $\mu_{rel}$ dimensions. \citet{Penny:2019} finds that the mean blend fraction of detected microlensing events is $\sim$ 0.2 meaning that the astrometric lensing signal could be diluted by a factor of $\sim$ 5. Regardless, our results demonstrates that accurate $\delta_{c, max}$ and $\mu_{rel}$ estimates will play an important role in separating PBH and BH events, making precise astrometric measurements with Roman a priority in the next decade of microlensing.

\section{Discussion and Conclusion}
\label{sec:Conclusion}

Dark matter makes up $\sim$85\% of the matter content in the universe, and though not well\hyp{}understood, may be predominately comprised of PBHs. Recent explorations have placed very tight constraints on $f_{\text{DM}}$ covering many decades in mass \citep[see summary in][]{Bird:2022}. Due to cosmological constraints, stellar evolved black holes can only account for a small portion of dark matter \citep{Planck:2014}, making detection and characterization of PBHs an important key in understanding the structure and formation of our universe. 

Throughout this work we have detailed and demonstrated the utility of adding a PBH population model into the microlensing simulation code \texttt{PopSyCLE}, enabling us to simulate microlensing surveys in the Milky Way, serving as a powerful tool for analyzing the estimated astrometric and photometric measurements of potential PBH microlensing events. Because this work is first and foremost a proof\hyp{}of\hyp{}concept for the addition of PBHs to \texttt{PopSyCLE} simulations, we make many simplifying assumptions about PBHs (see section \S\ref{sec:Primordial Black Hole Injection}) and microlensing survey parameters (see section \S\ref{sec:microlensing sims}). We acknowledge our assumption of $f_{\text{DM}}=1$ and monochromatic PBH mass of $\bar{m}_{\text{PBH}}=30$ $M_{\odot}$ do not necessarily align with recent constraints on $f_{\text{DM}}$ \cite[e.g.,][]{Blaineau:2022, 2001:Alcock, Tisserand:2007, Wyrzykowski:2011} or the expectation of an extended mass distribution, however they enable us to demonstrate our \texttt{PopSyCLE} + PBH simulations, and allows the reader to scale our results to other values of $f_{\text{DM}}$ and $\bar{m}_{\text{PBH}}$ as they see fit. Reported results scale roughly proportional to $f_{\text{DM}}$ and approximately constant as a function of mass, modulo potentially important survey selection effects.

Given our simplified PBH model and survey assumptions, we find the following:
\begin{itemize}
    \item Assuming a monochromatic PBH mass of $\bar{m}_{\text{PBH}}=30$ $M_{\odot}$, PBH microlensing events tend towards longer duration ($t_E$), smaller parallax ($\pi_E$), larger maximum astrometric shift ($\delta_{c, max}$), and larger source-lens proper motion ($\mu_{rel}$) than BHs, but in most cases PBHs overlap with the stellar BH population.
    \item \change{We predict Roman will have the ability to detect $\sim$ \maybe{4,981}$f_{\text{DM}}$ (\update{scaling method 1, less} conservative estimate, determined by area scaling ratios alone) PBH events, $\sim$ \maybe{1,195}$f_{\text{DM}}$ (\update{scaling method 2, more} conservative estimate, determined by area scaling ratios between our simulations and the Roman survey, along with the Roman duty cycle of 0.24)\update{, and} a more realistic estimate of $\sim$ \maybe{1,212}$f_{\text{DM}}$ events (\update{scaling method 3}, considering events that would fall above the 100 exposure precision threshold for \update{F146} source magnitude, and fall below the limiting magnitude of the survey, then scaling to the full survey area). These reported values assume that there is constant stellar density across Roman fields (allowing us to scale up to the full Roman survey area of 1.97 $\text{deg}^2$), \update{no survey gaps,} that $\bar{m}_{\text{PBH}}=30$ $M_{\odot}$ lies within a mass range detectable by Roman with moderate efficiency, \update{and that noise is in the source-dominated regime and neglects effects of blending}. We expect this to hold to the order of $\color{black}\sim$ \update{1,000}$f_{\text{DM}}$ after taking into account the multiple overestimations and underestimations described throughout this work, which are due to PBH population and survey assumptions.}
    \item \change{We find that $\sim$ 355 PBH events may be within the 8 year OGLE-IV dataset, in which each PBH may conceivably be detected, but reliable lensing parameter estimates may not be attainable for. We additionally estimate $\sim$ 306 PBH events may be detectable and also have sufficient SNR such that lensing estimates can be reasonably estimated for each PBH. We expect this to hold on the order of $10^2f_{\text{DM}}$.}
    \item With the assumption of $\bar{m}_{\text{PBH}}=30$, we find $\sim$ \change{4.3}$f_{\text{DM}}$ and $\sim$ \maybe{2.8}$f_{\text{DM}}$ times more PBH microlensing events than BH microlensing events for our Mock OGLE\hyp{}IV (EWS cut) and Roman simulations, respectively. The fraction of events is higher for the OGLE survey as it has a longer survey duration, and PBH microlensing events tend toward longer timescales. \change{This averages out to \maybe{3.6}$f_{\text{DM}}$ more PBHs than BHs.}
    \item Using simulated PBH distributions with $f_{\text{DM}}=1$ and $\bar{m}_{\text{PBH}}=30$, we find median $t_E$ values for our OGLE\hyp{}IV EWS cut and Roman simulations of $\sim$ 91.7 days and $\sim$ \maybe{91.3} days, respectively, where peak $t_E$  scales approximately proportional to $\sqrt{\bar{m}_{\text{PBH}}}$. \change{This results in an average PBH estimated peak Einstein crossing time of $\sim$ \maybe{91.5} days.}
\end{itemize}

Roman will have the ability to conduct simultaneous photometric and astrometric observations, with an astrometric sensitivity that enables significant astrometric lensing measurement for the majority of events where $M_{\text{PBH}} \geq 1$ $M_{\odot}$ (see Figure \ref{fig:pi_e_delta_c_dist_all}). Therefore, we expect Roman to be an important instrument in the next decade of microlensing science (and for potential PBH detection), in that it will enable better disambiguation between the PBH and BH event populations in the $\pi_E$, $t_E$, $\delta_{c, max}$ and $\mu_{rel}$ parameter spaces. Due to the overlap between PBHs and BHs in these various spaces, and our average result of $\sim$ \maybe{3.6}$f_{\text{DM}}$ times more 30 $M_{\odot}$ PBH microlensing events than BH events, surveying the Magellanic clouds may provide improved PBH dark matter fraction constraints, as the lower stellar density should result in a higher relative number of PBHs to BHs, enabling us to better constrain photometric and astrometric PBH parameters. To be able to simulate this, a Magellanic Cloud model would need to be added to \texttt{PopSyCLE}.

With all simplifying assumptions aside, we have demonstrated and validated our addition of a PBH model into \texttt{PopSyCLE} that can be used to analyze astrometric and photometric microlensing measurement predictions for PBHs, serving as a powerful tool in the next decade of microlensing science and astronomical surveys. 

\section{Acknowledgements} 
This work was performed under the auspices of the U.S. Department of Energy by Lawrence Livermore National Laboratory under Contract DE\hyp{}AC52\hyp{}07NA27344 and was supported by the LLNL\hyp{}LDRD Program under Project Numbers 17\hyp{}ERD\hyp{}120 and 22\hyp{}ERD\hyp{}037, and by LLNL HEP Program Development. M.S.M acknowledges support from the University of California Office of the President for the UC Laboratory Fees Research Program In\hyp{}Residence Graduate Fellowship (Grant ID: LGF\hyp{}19\hyp{}600357). C.Y.L. and J.R.L. acknowledge support by the National Science Foundation under grant No. 1909641, the National Aeronautics and Space Administration (NASA) under contract No. NNG16PJ26C issued through the WFIRST (now Roman) Science Investigation Teams Program, and the Heising\hyp{}Simons Foundation under Grant No. 2022\hyp{}3542. C.Y.L. also acknowledges support from NASA FINESST grant No. 80NSSC21K2043. LLNL\hyp{}JRNL\hyp{}834877.

\bibliography{references}

\end{document}